\documentclass[prb,a4paper,showpacs,twocolumn]{revtex4}

\usepackage{amssymb}
\usepackage{amsmath}
\usepackage{amsfonts}
\usepackage{graphicx}
\usepackage{color}
\usepackage{bm}

\renewcommand{\Re}{\,\textrm{Re}\,}
\renewcommand{\Im}{\,\textrm{Im}\,}

\DeclareMathOperator{\Sp}{sp} 
\DeclareMathOperator{\tr}{tr} 
 \DeclareMathOperator{\Imag}{Im}

\begin{document}

\title{Disordered electron liquid in double quantum well heterostructures: \\
Renormalization group analysis and dephasing rate}

\author{I.S.\,Burmistrov$^{1}$, I.V.\,Gornyi$^{2,3}$, and K.S.\,Tikhonov$^{4,1}$} 
\affiliation{$^{1}$L.D.\ Landau Institute for Theoretical Physics,
Russian Academy of Sciences, 117940 Moscow, Russia}
\affiliation{$^2$Institut f\"ur Nanotechnologie, Karlsruhe Institute of Technology, 76021 Karlsruhe, Germany}
\affiliation{$^{3}$A.F.Ioffe Physico-Technical Institute, 194021 St.Petersburg, Russia}
\affiliation{$^{4}$Department of Physics, Texas A\&M University, College Station, TX 77843, USA}

\begin{abstract}

We report a detailed study of the influence of the electron-electron interaction on physical observables (conductance, etc.) of 
a disordered electron liquid in double quantum well heterostructure. We find that even in the case of common elastic scattering off electrons in both quantum wells, the asymmetry in the electron-electron interaction across and within quantum wells decouples them at low temperatures. Our results are in quantitative agreement with recent transport experiments on the gated double quantum well Al$_x$Ga$_{1-x}$As/GaAs/Al$_x$Ga$_{1-x}$As heterostructures.

\end{abstract}
\date{\today}


\maketitle

%%%%%%%%%%%%%%%%%%%%%%%%%%%%%%%%%%%%%%%%%%%%%%%%%%%%%%%%%%%%%%%%%%%%%%%%%%%%%%%%

\section{Introduction \label{Sec_Intro}}

Disordered two-dimensional (2D) electronic systems have been remaining in the
focus of experimental and theoretical research for more than three
decades.~\cite{AFS} The experimental discovery~\cite{Pudalov1,prb95}  of the
metal-insulator transition (MIT) in a high mobility silicon
metal-oxide-semiconductor field-effect transistor
(Si-MOSFET) in 1994 became a challenge to a theory. Although during last decade
the behavior of resistivity similar to that of Ref.~[\onlinecite{Pudalov1,prb95}] has been found experimentally
in a wide variety of two-dimensional electron systems,~\cite{Review} the MIT in two dimensions still calls for deeper
theoretical and experimental understanding.

Very likely, the most promising theoretical framework for studying the 2D MIT is provided by the
effective low-energy theory, initially developed by Finkelstein, that
combines the diffusive dynamics due to disorder and strong electron-electron
interaction.~\cite{Finkelstein} 
Moreover, it is the Finkelstein theory that suggested metallic behavior at low temperatures 
long before the experimental discovery of the MIT in a Si-MOSFET.~\cite{Finkelstein,Castellani}
Recently, Punnoose and Finkelstein~\cite{LargeN} have shown a possibility for the existence of the MIT
in the special model of 2D electron system with $SU(\mathcal{N})$ degrees of freedom in the 
limit of the large number of multiplets, $\mathcal{N}\to \infty$. 
On the other hand, the current theoretical results~\cite{BPS,BBP} do not support the existence of MIT for 
electrons interacting in the singlet channel only ($\mathcal{N}=1$).
Therefore, the presence of additional degrees of freedom (spins, valley isospins etc.) plays a crucial 
role for the existence of the MIT in 2D disordered electron systems. 
In fact, the importance of the multiplet channels of the interaction 
has been confirmed experimentally in Si-MOSFET where a weak magnetic field 
applied parallel to the 2D plane changes the behavior of resistivity from metallic to insulating at low temperatures.~\cite{Pudalov2,Vitkalov,Pudalov2003} 
These experimental findings have been explained in framework of the Finkelstein theory in the presence of Zeeman and valley splitting.~\cite{BurmistrovChtchelkatchev} 
The effect of intervalley scattering has been taken into account as well.~\cite{Punnoose2}

Recently, the Finkelstein theory for disordered electron liquid in Si-MOSFET has been subjected to 
a detailed experimental check. In particular,  
the metallic behavior of resistivity not far away from the MIT,~\cite{FP}  
the increase of interaction parameter in the multiplet channels,~\cite{JETPL2, KravchenkoNew,Punnoose1} 
and the two-parameter scaling near MIT~\cite{PudalovPRL} have been observed in experiments.
Such the analysis in Si-MOSFET is complicated by the presence of (uncontrolled) large valley splitting 
and intervalley scattering rate, $\Delta_v\approx 1/\tau_v \approx 1 K$.~\cite{Kuntsevich,Klimov} 

As known,~\cite{Shayegan0} in  n-AlAs quantum
wells, 2D electrons can also populate two valleys. In addition to Si-MOSFETs this system offers opportunity for an
experimental investigation of the interplay between the spin and
valley degrees of freedom. Using a symmetry breaking strain to tune the valley occupation of the 2D electron system in
the n-AlAs quantum well, as well as a parallel magnetic field to
adjust the spin polarization, the spin -- valley interplay has been
experimentally studied.~\cite{Shayegan1,Shayegan2} However, the
electron concentrations in the experiment were at least three
times larger than the critical one corresponding to the MIT.~\cite{Shayegan0} Therefore,
the spin-valley interplay in n-AlAs quantum well has been studied only in the region of a
good metal, very far from the MIT.

Disordered electron liquid in double quantum well heterostructures represents a 2D system 
in which electrons in addition to spin have the other degree of freedom: the isospin associated with a quantum well. 
In spite of a number of interesting physical phenomena observed in electron liquids in
double quantum well heterostructures without and under strong magnetic field, e.g., Coulomb drag,~\cite{DragExp} Bose-Einstein condensation of excitons~\cite{Current}, ferromagnetic~\cite{Fer} and canted antiferromagnetic phases,~\cite{AFer} the metal-insulator transition has not been yet addressed experimentally.
Transport of electrons in double quantum well heterostuctures has been studied experimentally~\cite{Portal,MinkovOld} only in the metallic regime far from the region in which MIT is expected.

Recently, detailed experimental research on the interference and interaction corrections to conductance 
of electrons in a double quantum well heterostructure has been performed.~\cite{MinkovGermanenko1,MinkovGermanenko2} 
In particular, two very distinct physical situations have been investigated: i) both quantum wells have equal electron concentrations and mobilities; ii) one quantum well remains with almost the same electron concentration as in case i), whereas the other is empty by applying the gate voltage.  Surprisingly, it was found that the 
dephasing rate and interaction correction to the conductance are almost the same for cases i) and ii).

In the present paper, motivated by the experiments of Refs.~[\onlinecite{MinkovGermanenko1,MinkovGermanenko2}],  
we develop the theory of the disordered electron liquid formed in a heterostructure with two almost identical quantum wells. 
We concentrate on the case of equal electron concentrations and mobilities in both quantum wells [corresponding to the case i) of 
Refs.~\onlinecite{MinkovGermanenko1,MinkovGermanenko2}]. This case will be termed as balance in what follows. 

We restrict our study to temperatures ($T$) satisfying the following condition: $1/\tau_{+-}, \Delta_s,\Delta_{SAS} \ll T \ll 1/\tau_{\rm tr}$. Here $1/\tau_{+-}$ stands for the rate of elastic scattering between symmetric and antisymmetric states in the double quantum well structure,  $\Delta_{SAS}$ the splitting of these symmetric and antisymmetric states, $\Delta_s$ the Zeeman splitting, and  $\tau_{\rm tr}$  the elastic transport mean free time. The temperature behavior of the interaction correction to the total conductance is governed by one singlet and $15$ multiplet diffusive modes. We find that the latter splits into three inequivalent groups of one, six, and eight modes. This grouping occurs due to asymmetry in electron-electron interactions across and within quantum wells which breaks the rotational symmetry in the combined spin and isospin spaces [$SU(4)$]. This reduced symmetry is a distinctive feature of double quantum well heterostructures at the balance and is absent in two-valley systems in Si-MOSFETs and n-AlAs quantum wells. We identify all relevant interaction parameters and estimate their dependence on the distance between the quantum wells. To describe the system at low temperatures and beyond interaction corrections to conductance, we derive the non-linear sigma model and study its renormalization in the one-loop approximation. As we demonstrate, the renormalization group equations describing the length scale dependence of the total conductance and interaction parameters drive the system towards the fixed point corresponding to two separate quatum wells. In spite of the symmetry breaking between $15$ multiplet modes, the renormalization group equations predict the metallic behavior of the conductance at low temperatures. 
Finally, we generalize the expression for the dephasing rate of electrons due to the presence of electron-electron interaction known~\cite{Aronov-Altshuler,Narozhny} for a single quantum well to the case of double quantum well heterostructures. We find that our results are in good quantitative agreement with experimental data of Refs.~[\onlinecite{MinkovGermanenko1,MinkovGermanenko2}].

The paper is organized as follows. In Section~\ref{Sec_Form}
we introduce the microscopic Hamiltonian, identify relevant interaction parameters, study its dependence on the distance between quantum wells and introduce the nonlinear sigma model that describes the
low-energy excitations in the disordered interacting electron system. Then, in Sec.~\ref{Sec_RG} we consider the renormalization of the nonlinear sigma model in the one-loop approximation, derive corresponding renormalization group equations, and discuss renormalization group flow. We derive expressions for the dephasing rate due to electron-electron interactions in Sec.~\ref{Sec_Deph}. 
Next in Sec.~\ref{Sec_Disc} we perform detailed comparison between our theory and recent experimental data on transport in double quantum well heterostructures. We end the paper with conclusions (Sec.~\ref{Sec_Conc}).

%%%%%%%%%%%%%%%%%%%%%%%%%%%%%%%%%%%%%%%

\section{Formalism\label{Sec_Form}}

\subsection{Microscopic Hamiltonian\label{Sec_Form_Ham}}

We consider 2D interacting electrons in double quantum well heterostructures in the presence
of quenched disorder at low temperatures $T\ll \tau _{\rm tr}^{-1}$. In the case of two almost identical quantum wells
an electron annihilation operator can be written as a linear combination of symmetric and antisymmetric states:
\begin{equation}
\psi^{\sigma}(\bm{R}) = \psi_{\tau }^{\sigma}(\bm{r}) \varphi_\tau(z),\quad
\varphi_\tau(z)= \frac{\varphi_l(z)+\tau \varphi_r(z)}{\sqrt{2}} .\label{WF}
\end{equation}
Here electron motion along $z$ axis is confined by the quantum wells, $\bm{r}$ denotes a vector in plane perpendicular to the $z$ axis, and $\bm{R} = \bm{r}+z \bm{e_z}$. 
The superscript $\sigma=\pm$ denotes electron spin projection,
$\tau=\pm$ enumerates symmetric ($+$) and antisymmetric ($-$) states in the double quantum well structure and $\psi^\sigma_\tau$ is the annihilation operator of an electron with the spin and isospin projections equal $\sigma/2$
and $\tau/2$, respectively. The normalized envelope function
$\varphi_{l,r}(z)=\varphi(z\pm d/2)$
corresponds to the wave function of an electron localized in a single left/right well.
In what follows, we assume a negligible overlap between the states in two quantum wells: 
the width of an electron state in a quantum well  $[\int dz\, \varphi^4(z)]^{-1} \ll d$ where $d$ is the distance between the centers of the quantum wells.

In the path-integral formulation, interacting
electrons in the presence of the random potential $V(\bm{R})$ 
are described by the following grand partition function
\begin{equation}
Z = \int \mathcal{D}[\bar{\psi},\psi] e^{{S}[\bar{\psi},\psi]}
\end{equation}
with the imaginary time action ($\beta=1/T$)
\begin{equation}
{S} = -\int_0^{\beta}\!\! dt \Bigl
\{ \int d\bm{r}  \bar{\psi}^{\sigma}_{\tau}(\bm{r}t)\bigl [ 
\partial_t+\mathcal{H}_0\bigr ] {\psi}^{\sigma}_{\tau}(\bm{r}t) 
-\mathcal{L}_\textrm{dis}-\mathcal{L}_\textrm{int}\Bigr
\}. \label{Hstart}
\end{equation}
The single-particle Hamiltonian 
\begin{gather}
\mathcal{H}_0 =
-\frac{\nabla^2}{2m_e}-\mu + \frac{1}{2} (\Delta_s\sigma +
\Delta_{\rm SAS}\tau)\label{Hstart0}
\end{gather}
describes a 2D quasiparticle with mass $m_e$. Magnetic field $B$ perpendicular to the $z$ axis induces 
the Zeeman splitting $\Delta_s=g \mu_B B$. The energy difference between symmetric and antisymmetric states in a double quantum well structure yields the splitting $\Delta_{SAS} \simeq 2 \varphi(d/2)\varphi^\prime(d/2)/m_e$.~\cite{LL3}  The chemical potential is denoted as $\mu$, $g$ stands for the effective electron $g$-factor and $\mu_B$ the Bohr magneton. The single-particle Hamiltonian~\eqref{Hstart0} is completely analogous to one for a Si(001)-MOSFET. In latter case, index $\tau$ enumerates valleys and $\Delta_{SAS}$ plays a role of a valley splitting. 

%The splitting $\Delta_{SAS}$ is exponentially small in  the large parameter $d\int dz\, \varphi_{l,r}^4(z)]$.  For example, in the double quantum well %heterostructures of Ref.~\cite{MinkovGermanenko1} $\Delta_{SAS}$ was estimated as $1\,K$.  

Next,the term 
\begin{equation}
\mathcal{L}_\textrm{dis} =-\int d \bm{r}\,
\bar{\psi}^{\sigma}_{\tau_1}(\bm{r}t)
V_{\tau_1\tau_2}(\bm{r}) {\psi}^{\sigma}_{\tau_2}(\bm{r}t)
\end{equation}
describes electron scattering off a random potential $V(\bm{R})$. 
It involves matrix elements
\begin{gather}
V_{\tau_1\tau_2}(\bm{r}) = \int dz\,
V(\bm{R}) \varphi_{\tau_1}(z)\varphi_{\tau_2}(z)
. \label{Veqr}
\end{gather}
In general, the matrix elements $V_{\tau_1\tau_2}$ induce transitions between symmetric and antisymmetric states in a double quantum well structure. In the case of symmetric random potential: $V(\bm{r},z) = V(\bm{r},-z)$, the system is protected from the symmetric-antisymmetric scattering. 

In accordance with the experimental conditions reported in Ref.~[\onlinecite{MinkovGermanenko1,MinkovGermanenko2}],  
we assume that impurities are concentrated in the middle between two quantum wells.  
We suppose that the random potential created by impurities has the Gaussian distribution, and
\begin{equation}
\langle V(\bm{R})\rangle = 0,\,\, \langle
V(\bm{R}_1)V(\bm{R}_2)\rangle =
W(|\bm{r}_1-\bm{r}_2|,|z_1|,|
z_2|),
\end{equation}
where $W$ decays as the function of its variables at a typical distance $d_W$. 
If the condition 
\begin{equation}
\Bigl [\int dz\, \varphi^4(z)\Bigr ]^{-1} \ll d, \label{Cond1}
\end{equation}
holds, we can neglect the small difference [proportional to $\varphi(d/2)\varphi^\prime(d/2)$], between symmetric-symmetric and antisymmetric-antisymmetric scattering rates. Then
\begin{gather}
\langle
V_{\tau_1\tau_2}(\bm{r}_1)V_{\tau_3\tau_4}(\bm{r}_2)\rangle
=W\left (|\bm{r}_1-\bm{r}_2|,d/2,d/2\right ) \delta_{\tau_1\tau_2}\delta_{\tau_3\tau_4}.
\label{W2}
\end{gather}
Provided correlations in $W$ are short-ranged,~\cite{Footnote} we find
\begin{gather}
\langle
V_{\tau_1\tau_2}(\bm{r}_1)V_{\tau_3\tau_4}(\bm{r}_2)\rangle
= \frac{1}{2\pi\nu\tau_i}
\delta_{\tau_1\tau_2}\delta_{\tau_3\tau_4}\delta(\bm{r}_1-\bm{r}_2),
\label{RandPot1}
\\
\frac{1}{\tau_i}=2\pi \nu \int d^2\bm{r}\,
W(|\bm{r}|,d/2,d/2) . \notag
\end{gather}
Here $\nu$ is the thermodynamic density of states of 2D electrons (including spin). We emphasize that 
electrons in both quantum wells are subjected to correlated disorder 
since they scatter off the very same random potential. 
Recently, under such assumptions, the transconductance of a double quantum well structure (the Coulomb drag effect with correlated disorder)  has been studied by one of the authors.~\cite{DragIgor}

The small asymmetry in the impurity distribution along $z$ axis will lead to the scattering between symmetric and antisymmetric states in the double quantum well structure. Its rate can be estimated as $1/\tau_{+-} \sim (b/d)^2/\tau_i \ll 1/\tau_i$ where $b$ is a typical length characterizing asymmetry. We neglect $1/\tau_{+-}$ in what follows.  
 
The interaction part of the action~\eqref{Hstart} reads
\begin{equation}
\mathcal{L}_{\rm int}=-\frac{1}{2}\int d \bm{R} d\bm{R^\prime} \rho(\bm{R}t)\, U(|\bm{R}-\bm{R^\prime}|)\, \rho(\bm{R^\prime}t) 
\end{equation}
where $U(\bm{R}) = e^2/\epsilon R$. The dielectric constant is denoted as $\epsilon$.
Expanding the density operator $\rho(\bm{R}t)=\bar{\psi}^{\sigma}_{\tau_{1}}(\bm{r}t)
\psi^{\sigma}_{\tau _{2}}(\bm{r}t)\varphi_{\tau _{1}}(z)\varphi _{\tau _{2}}(z)$
and assuming again that condition~\eqref{Cond1} holds we obtain
\begin{gather}
\mathcal{L}_{\rm int}=-\frac{1}{8}\int d\bm{r} d\bm{r^\prime}\, 
\bar{\psi}^{\sigma_1}_{\tau_{1}}(\bm{r}t) \psi^{\sigma_1}_{\tau _{2}}(\bm{r}t)
\bar{\psi}^{\sigma_2}_{\tau_{3}}(\bm{r^\prime}t) \psi^{\sigma_2}_{\tau _{4}}(\bm{r^\prime}t)
\notag \\
\times 
\Bigl [ (1
+\tau_{1}\tau_{2}\tau_{3}\tau_{4}) U_{11}(|\bm{r}-\bm{r^\prime}|) \notag \\ 
+
(\tau_{1}\tau_{2}+\tau_{3}\tau_{4}) U_{12}(|\bm{r}-\bm{r^\prime}|) \Bigr] .
\end{gather}
Here 
\begin{equation}
U_{11}(r) =\frac{e^2}{\epsilon} \int dz dz^\prime \frac{\varphi_l^{2}(z) \varphi_l^{2}(z^\prime)}{\sqrt{r^2
+(z-z^\prime)^2}} \approx \frac{e^2}{\epsilon r}
\end{equation}
is the standard Coulomb interaction between electrons in a single well. The interaction between electrons in different quantum wells
\begin{equation}
U_{12}(r) =\frac{e^2}{\epsilon} \int dz dz^\prime \frac{\varphi_l^{2}(z) \varphi_r^{2}(z^\prime)}{\sqrt{r^2
+(z-z^\prime)^2}} \approx \frac{e^2}{\epsilon \sqrt{r^2+d^2}}
\end{equation}
takes into account that electrons are separated by the distance $d$. Due to the difference between $U_{11}$ and $U_{12}$ the interaction Lagrangian $\mathcal{L}_{\rm int}$ is
not invariant under global $SU(4)$ rotations of the
electron operator $\psi^\sigma_\tau$ in the combined spin-isospin space. It is the interaction part of the action~\eqref{Hstart} that distinguishes the disordered electron liquid in double quantum well heterostructures from the one in a Si(001)-MOSFET.

As usual, we single out regions in the momentum space of small momentum transfer~\cite{Finkelstein,Castellani,KirkpatricBelitz,AleinerZalaNarozhny}. 
Then the low energy part of $\mathcal{L}_\textrm{int}$ can be written as
\begin{gather}
\mathcal{L}_\textrm{int} = \frac{1}{4\nu} \int^\prime \frac{d\bm{q}}{(2\pi)^2}
\sum_{a,b=0}^3  \mathbb{F}_{ab}(q) m^{ab}(\bm{q}) 
m^{ab}(-\bm{q}) ,\\
m^{ab}(\bm{q}) =  \int \frac{d\bm{k}}{(2\pi)^2} \bm{\bar\psi}(\bm{k}+\bm{q}) t_{ab} \bm{\psi}(\bm{k}) .
\end{gather}
Here $\bm{\bar\psi} = \{\bar\psi^+_+,\bar\psi^-_+,\bar\psi^+_-,\bar\psi^-_-\}$,  $\bm{\psi} = \{\psi^+_+,\psi^-_+,\psi^+_-,\psi^-_-\}^T$, the `prime' at the integral sign denotes the integration region $q\lesssim l^{-1}$ ($l$ is the elastic mean free path), and 16 matrices 
$t_{ab} = \bm{\tau}_a \otimes \bm{\sigma}_b$ stand for the generators of $SU(4)$. 
Pauli matrices $\bm{\tau}_a$, $a=0,1,2,3$
act in the isospin space of two wells and Pauli matrices $\bm{\sigma}_b$, $b=0,1,2,3$ act in the spin space. The matrix of interaction parameters reads
\begin{equation}
\mathbb{F}(q) = \begin{pmatrix}
F_s & F_t &F_t&F_t \\
\tilde F_s & F_t &F_t&F_t \\
F_v & F_v &F_v &F_v \\
F_v & F_v &F_v &F_v 
\end{pmatrix}\label{Fmatrix}
\end{equation}
where
\begin{gather}
F_t =  - \frac{\nu}{2} \langle U_{11}^{\rm scr}(0)\rangle_{FS},\qquad 
F_v =  -\frac{\nu}{2} \langle U_{12}^{\rm scr}(0)\rangle_{FS} ,\notag \\ 
F_s = \nu [U_{11}(q) + U_{12}(q)] +F_t , \notag \\
\tilde{F}_s = \nu[U_{11}(q) - U_{12}(q)]+F_t . \label{FFs} 
\end{gather}
Here $U_{11}(q) = 2\pi e^2/q\epsilon$, $U_{12}(q) = U_{11}(q) \exp(-qd)$. The quantities $F_t$ and $F_v$ are analogous 
to the standard Fermi liquid interaction parameters in the triplet channel. 
They involve averaging of the static part of dynamically screened
interaction $U_{11/12}^{\rm scr}(q,\omega)$ over the Fermi surface. 
In the case of equal electron concentrations and mobilities in both quantum wells
\begin{equation}
 \langle U_{11/12}^{\rm scr}(0)\rangle_{FS} = \int_0^{2\pi} \frac{d\theta}{2\pi} U_{11/12}^{\rm scr}(2k_F\sin(\theta/2),0) 
\end{equation}
where $k_F$ is Fermi momentum for a single quantum well. The interaction parameter $F_s$ involves the
long-range part of the Coulomb interaction. In the limit $q\to 0$ it becomes $F_s(q)\approx 2\varkappa/q \to \infty$  
where $\varkappa = 2\pi e^2 \nu/\epsilon$.
Within the same accuracy, we find 
\begin{equation}
\tilde{F}_s = \varkappa d + F_t . \label{Fst}
\end{equation}

At $d=0$---when both quantum wells coincide---the interaction parameters are equal: $\tilde{F}_s=F_t=F_v$. Then the matrix $\mathbb{F}$ corresponds to the case of electron liquid with two valleys as it occurs in Si(001)-MOSFET. In the absence of $\Delta_{SAS}$ and $\Delta_s$ the action~\eqref{Hstart} becomes invariant under global $SU(4)$ rotations of the fermionic fields. In the opposite case of 
$d\to \infty$, the double quantum well heterostructure is equivalent to two independent single quantum wells. Then we obtain 
$\tilde F_s=F_s$, and $F_v=0$. The action~\eqref{Hstart} (for $\Delta_{SAS}=\Delta_s=0$) becomes invariant under global $SU(2)$ rotations of electron spin in each quantum well independently. For intermediate values of $d$, the action~\eqref{Hstart} is also invariant under global $SU(2)\times SU(2)$ rotations provided $\Delta_{SAS}$ and $\Delta_s$ vanish.

%%%%%%%%%%%%%%%%%%%%%%%%%%%%%%%%%%%%%%%%%%%%%%%%%%%%%%%%%%%%%%%%%%%%%%%%%%%%%
\begin{figure}[t]
\centerline{\includegraphics[width=80mm]{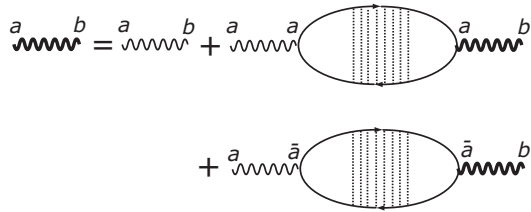}}
\caption{Dyson equation for the screened electron-electron interaction in RPA. Thick wavy line denotes screened interaction, thin wavy line is bare interaction, solid lines are electron Green's functions, and dashed lines are impurity lines. Indices $a$ and $b$ can be $1$ or $2$. Index $\bar a$ equals $1$ ($2)$ if index $a$ is $2$ ($1$). }
\label{RPA_Figure}
%\vspace{-0.5cm}
\end{figure}
%%%%%%%%%%%%%%%%%%%%%%%%%%%%%%%%%%%%%%%%%%%%%%%%%%%%%%%%%%%%%%%%%%%%%%%%%%%

\subsection{Dynamically screened Coulomb interaction\label{Sec_Form_DynScr}}

The interaction parameters $F_t$ and $F_v$ involve the screened Coulomb interaction. Solving the Dyson equations in the random phase approximation (RPA) (see Fig.~\ref{RPA_Figure}), we obtain the following results for the dynamically screened interactions:~\cite{KamenevOreg,Flensberg}
\begin{gather}
U_{11}^{\rm scr} = \frac{U_{11} + \Pi_2[U_{11}^2 - U_{12}^2] }{1+[\Pi_1+\Pi_2]U_{11}+ \Pi_1\Pi_2[U_{11}^2 - U_{12}^2]} ,
\label{ScrU11}\\
U_{12}^{\rm scr} = \frac{U_{12}}{1+[\Pi_1+\Pi_2]U_{11}+ \Pi_1\Pi_2[U_{11}^2 - U_{12}^2]}, \label{ScrU12}  \\
U_{22}^{\rm scr} = \frac{U_{11} + \Pi_1[U_{11}^2 - U_{12}^2] }{1+[\Pi_1+\Pi_2]U_{11}+ \Pi_1\Pi_2[U_{11}^2 - U_{12}^2]} .
\label{ScrU22}
\end{gather}
The polarization operators can be written in diffusive approximation as
\begin{equation}
\Pi_j(q,\omega) = \nu \frac{D_j q^2}{D_j q^2-i\omega},\quad j=1,2
\end{equation}
where $D_j$ is the diffusion coefficient in the $j$-th quantum well. 
We mention that for $D_1\neq D_2$ the dynamically screened Coulomb interaction in the first well $U_{11}^{\rm scr}(q,\omega)$ does not coincide with the one ($U_{22}^{\rm scr}(q,\omega)$) in the second well. 

 If the electron concentrations and mobilities in the quantum wells are the same then $D_1=D_2$. 
In this case $U_{11}^{\rm scr}=U_{22}^{\rm scr}$ and 
\begin{gather}
U_{11/12}^{\rm scr}  = \frac{\varkappa}{2 \nu q} \Bigl (Dq^2-i\omega\Bigr ) \Biggl \{ \frac{1+e^{-qd}}{Dq\Bigl [q +\varkappa(1+e^{-qd})\Bigr]-i\omega} \notag \\
\pm 
\frac{1-e^{-qd}}{Dq\Bigl [q +\varkappa(1-e^{-qd})\Bigr]-i\omega} \Biggr \} .
\end{gather}
As one can see, at $qd\gg 1$ the effect of the right well on the dynamically screened interaction in the left well is negligible. In the opposite case, $qd\ll 1$ the right well affects the dynamically screened interaction in the left well only at $\varkappa d\lesssim 1$.

\subsection{Estimates for the interaction parameters \label{Sec_Form_IntPar}}

\begin{figure}[t]
\centerline{\includegraphics[width=75mm]{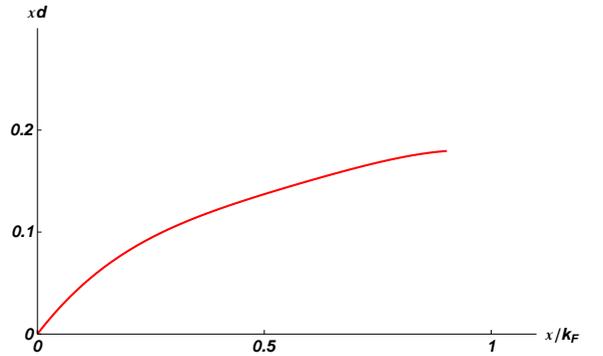}}
\caption{Value of the parameter $\varkappa d$ at which $\tilde F_s=0$ versus $\varkappa/k_F$.}
\label{CriticalD_Figure}
%\vspace{-0.5cm}
\end{figure}

Let us estimate the interaction parameters $F_t$ and $F_v$ in the case of equal electron concentrations in both quantum wells. By using Eqs.~\eqref{ScrU11} and \eqref{ScrU12} we find 
\begin{equation}
F_t\pm F_v = -\int\limits_0^{2\pi} \frac{d\theta}{4\pi} \frac{\varkappa (1\pm e^{-2k_Fd\sin\theta/2})}{2k_F\sin\frac{\theta}{2}+\varkappa  (1\pm e^{-2k_Fd\sin\theta/2})} . \label{FtvPM}
\end{equation}
To justify the RPA which has been used in derivation of Eqs.~\eqref{ScrU11}-\eqref{ScrU12} we assume that the condition $\varkappa/k_F\ll 1$ holds.  
As follows from Eq.~\eqref{FtvPM}, both $F_t$ and $F_v$ are negative and $|F_t| \geqslant |F_v|$.
The interaction parameter $\tilde F_s$ is negative at small $d$ and positive at large $d$. The dependence of the critical distance $d_c$ at which $\tilde F_s$ vanishes on the parameter $\varkappa/k_F$ is shown in Fig.~\ref{CriticalD_Figure}. We mention that $|\tilde F_s|\leqslant |F_t|$
for $d<d_c$.
%, the absolute value of the interaction parameter $|\tilde F_s|\leqslant |F_t|$. 

It is instructive to compare the results for $F_t$, $F_v$ and $\tilde F_s$ with the case of a single quantum well for which the interaction parameter in the triplet channel is given as~\cite{AleinerZalaNarozhny}
\begin{gather}
F_t^0 = -\int\limits_0^{2\pi} \frac{d\theta}{4\pi} \frac{\varkappa }{2k_F\sin(\theta/2)+\varkappa} 
%=  - \int\limits_0^{\pi/2} \frac{d\theta}{\pi} \frac{\varkappa }{2k_F\sin(\theta)+\varkappa}  = - \int\limits_0^{1} \frac{d u}{\pi} \frac{\varkappa }{4k_F u+\varkappa(1+u^2)} 
= -\frac{1}{2\pi} \mathcal{G}_0(\varkappa/2k_F) ,\notag \\
\mathcal{G}_0(x) = \frac{x}{\sqrt{1-x^2}} \ln \frac{1+\sqrt{1-x^2}}{1-\sqrt{1-x^2}} .\label{Ft0}
\end{gather}
In the limit $x\to 0$ the function $\mathcal{G}_0(x)$ acquires the following asymptotic form
\begin{equation}
\mathcal{G}_0(x) \approx x \ln (2/x),\qquad x\ll 1 .
\end{equation}
Provided $k_F d\gg 1$, the interaction parameters for the case of double quantum wells with equal electron concentrations can be estimated as
\begin{gather}
F_t = F_t^0 +\frac{1}{8\pi k_F d} \mathcal{G}_1(\varkappa d), \quad F_v = \frac{1}{8\pi k_F d} \mathcal{G}_2(\varkappa d) ,\label{Ftv} \\ 
\mathcal{G}_1(x) = \frac{3 x\, e^x E_1(x)}{x+1} + \frac{2x\, e^{-2x/(x-1)}}{x^2-1} 
E_1\left (-\frac{2x}{x-1}\right ) , \notag \\ 
\mathcal{G}_2(x) = \mathcal{G}_1(x)-\frac{4 x\, e^x E_1(x)}{x+1}  .
\notag 
\end{gather}
Here $E_1(x) = \int_x^\infty dt\, \exp(-t)/t$ is the exponential integral.  

Finally, we mention that the interaction parameters $F_t$ and $F_v$ can be estimated (from above) as
$|F_t| \leqslant [\mathcal{G}_0(\varkappa/k_F)+\mathcal{G}_0(\varkappa/2k_F)]/(4\pi)$ and $|F_v| \leqslant \mathcal{G}_0(\varkappa/k_F)/(4\pi)$. Even for values of $\varkappa/k_F \sim 1$, it yields $|F_t| \lesssim 0.3$ and $|F_v| \lesssim 0.2$.

%%%%%%%%%%%%%%%%%%%%%%%%%%%%%%%%%%%%%%%%%%%%%%%%%%%%%%%%%%%%%%%%%%%%%%%%%%%%%
%%%%%%%%%%%%%%%%%%%%%%%%%%%%%%%%%%%%%%%%%%%%%%%%%%%%%%%%%%%%%%%%%%%%%%%%%%%

\subsection{Non-linear $\sigma$ model\label{Sec_Form_NLSM}}

At low temperatures, $T\tau_\textrm{tr}\ll 1$, the effective
quantum theory of 2D disordered interacting electrons described by
the microscopic action~\eqref{Hstart} is given in terms of the non-linear
sigma model.  The latter describes interaction between low-enegy modes which are  the
so-called ``Diffusons'' and ``Cooperons''. As well-known,~\cite{Aronov-Altshuler,AALK81,AA81} the interference (``Cooperon'') contribution to the conductance is not sensitive to the presence of $\Delta_s$ and $\Delta_{SAS}$ (in the absence of $1/\tau_{+-}$). 
Furthermore, the interference correction is cut off by weak magnetic fields and does not influence the scaling of observables with temperature at $B \gtrsim 1/eD\tau_\varphi$.  Therefore, we shall ignore the interference correction in the intermediate calculations for a sake of simplicity and shall discuss its role in Sec.~\ref{Sec_Disc}.

In general, Cooperons are also involved in the interaction correction to the conductance and the renormalization of other interaction couplings. The corresponding contributions are proportional to the interaction parameter in the Cooper channel. For Coulomb interaction, the latter is repulsive and remains small in the course of the renormalization for 2D electron systems.~\cite{Finkelstein} 
Moreover, physically, a moderately weak magnetic field $B \gtrsim T/eD$ applied parallel to the $z$ axis is enough to suppress the interaction effects in the Cooper channel.~\cite{ALKL} 

Neglecting the Cooper channel,
the effective theory involves unitary matrix field
variables $Q^{\alpha_1\alpha_2;\sigma_1\sigma_2}_{mn;\tau_1\tau_2}(\bm{r})$
which obey the nonlinear constraint $Q^2(\bm{r})=1$. The
integers $\alpha_j=1,2,\dots,N_r$ denote the replica indices. The
integers $m, n$ correspond to the discrete set of Matsubara
frequencies $\varepsilon_n=\pi T (2n+1)$. 

The
effective sigma-model action is
\begin{equation}
\mathcal{S} = \mathcal{S}_\sigma +\mathcal{S}_F+
\mathcal{S}_{SB}.
\label{Sstart}
\end{equation}
Here $\mathcal{S}_\sigma$ represents the free electron
part~\cite{FreeElectrons}
\begin{equation}
\mathcal{S}_\sigma = -\frac{\sigma_{xx}}{32} \int d\bm{r}\tr (\nabla Q)^2
\label{SstartSigma}
\end{equation}
with $\sigma_{xx}=4\pi\nu_* D$ denoting the mean-field conductance in units of
$e^2/h$. The thermodynamic density of states $\nu_*=m_*/\pi$  involves an effective
mass $m_*$ renormalized due to interactions. 
The symbol $\tr$ stands for the trace over replica, the
Matsubara frequencies, spin and isospin indices. The Finkelstein term~\cite{Finkelstein,Unify}
\begin{gather}
\mathcal{S}_F = - \frac{\pi T}{4} \int d \bm{r}\sum_{\alpha n; a b} \bm{\Gamma}_{ab} \tr
I_n^\alpha t_{ab} Q(\bm{r}) \tr I_{-n}^\alpha t_{ab} Q(\bm{r}) \notag \\
+4\pi T z  \int d \bm{r}\tr \eta (Q-\Lambda) -2\pi T z  \int d \bm{r} \tr\eta\Lambda
 \label{SstartF}
\end{gather}
involves the electron-electron interaction amplitudes $\bm{\Gamma}_{ab}$. 
The bare value of the factor $z$ is determined by the thermodynamic density of states: $z= \pi \nu_*/4$. 
The quantity $z$ has been originally introduced by
Finkelstein in order to ensure the consistence of the renormalization group equations with the particle number conservation.~\cite{Finkelstein}  
Physically, the renormalization of $z$ is responsible for renormalization of the
specific heat~\cite{CasDiCas} and determines the relation between the frequency and length scales,
thus playing a crucial role at the criticality near the MIT.~\cite{Finkelstein} 

The interaction
amplitudes $\bm{\Gamma}_{ab}$ are related to the interaction parameters $\mathbb{F}_{ab}$ introduced above as~\cite{Finkelstein,Castellani,KirkpatricBelitz} $\bm{\Gamma}_{ab} = -z \mathbb{F}_{ab}/(1+\mathbb{F}_{ab})$. Therefore, the matrix $\bm{\Gamma}$ has the structure similar to the matrix $\mathbb{F}$ (see Eq.~\eqref{Fmatrix}) and 
\begin{gather}
\Gamma_s = -z,\,\, \tilde \Gamma_s = - z \tilde\gamma_s, \,\, \Gamma_t = - z \gamma_t, \,\, \Gamma_v = - z \gamma_v, \notag \\
\tilde\gamma_s = -\frac{\tilde F_s}{1+\tilde F_s},\,\, \gamma_t = -\frac{F_t}{1+F_t},\,\, \gamma_v=-\frac{F_v}{1+F_v} .\,
\end{gather}
The matrices
$\Lambda$, $\eta$ and $I_k^\gamma$ are given as
\begin{gather}
\Lambda^{\alpha\beta}_{nm} =
\mathrm{sign}\,(\omega_n)
\delta_{nm}\delta^{\alpha\beta} t_{00},\qquad 
\eta^{\alpha\beta}_{nm} = n
\delta_{nm}\delta^{\alpha\beta}t_{00},
\notag \\
(I_k^\gamma)^{\alpha\beta}_{nm} =
\delta_{n-m,k}\delta^{\alpha\gamma}\delta^{\beta\gamma}t_{00}.\label{matrices_def}
\end{gather}

The action $\mathcal{S}_\sigma+\mathcal{S}_F$ is invariant under the global
rotations $Q_{nm;\tau_1\tau_2}^{\alpha\beta;\sigma_1\sigma_2}(\bm{r})
\to u_{\sigma_1\sigma_3}^{\tau_1\tau_3}
Q_{nm;\tau_3\tau_4}^{\alpha\beta;\sigma_3\sigma_4}(\bm{r})
[u^{-1}]_{\sigma_4\sigma_2}^{\tau_4\tau_2}$ with $u = \sum_{a=0}^1\sum_{b=0}^3 u_{ab} t_{ab}$. 
This rotation correspond to the global $SU(2)\times SU(2)$ symmetry of the action $\mathcal{S}_\sigma+\mathcal{S}_F$. 

The presence of $\Delta_s$ and/or $\Delta_{SAS}$ generates the symmetry
breaking terms. In general, they can be written as~\cite{Finkelstein}
\begin{equation}
\mathcal{S}_{SB}=  i z_{ab} \Delta_{ab} \int
d\bm{r} \tr t_{ab} Q + \frac{N_r z_{ab} }{\pi T} \int
d\bm{r} \Delta_{ab}^2 . \label{Sstartsb}
\end{equation}
For the symmetry breaking by the Zeeman splitting one can choose $t_{ab} = t_{03}$ and $\Delta_{03}=\Delta_s$. In the case of the splitting $\Delta_{SAS}$, the generator $t_{ab}$ equals $t_{30}$. Splitting $\Delta_{ab}$ set the cut-off for a pole in the diffusion
modes (``diffusons''). In what follows, we shall be interested in high temperatures ($T \gg \Delta_{ab}$) or, correspondingly, in short length scales $L\ll \sqrt{D/\Delta_{ab}}$ such that the cut-off is irrelevant and the electron system behaves as if no symmetry breaking terms are
exist. We shall use the symmetry breaking term $\mathcal{S}_{SB}$ only as a source, assuming infinitesimal $\Delta_{ab}$.

%The spin splitting $\Delta_s$ and $\Delta_{SAS}$ set the cut-off for a pole in the diffusion
%modes (``diffusons'') with opposite spin and isospin
%projections. In what follows, we shall be interested in high temperatures ($T \gg \Delta_{SAS}, \Delta_s$) or, correspondingly, short length scales $L\ll \sqrt{D/\Delta_{SAS}}, \sqrt{D/\Delta_s}$ such that the cut-off is irrelevant and the electron system behaves as if no symmetry breaking terms are
%exist. We shall use the symmetry breaking term $S_{SB}$ only as a source assuming infinitesimal $\Delta_{ab}$. 

\subsubsection{$\mathcal{F}$-algebra \label{Sec:Formalism:Finv}}

The action~\eqref{Sstart} involves the matrices which are formally
defined in the infinite Matsubara frequency space. In order to
operate with them we have to introduce a cut-off for the Matsubara
frequencies. One should send the cut-off to infinity at the end of all calculations. 
Then, the set of rules which is called
$\mathcal{F}$-algebra can be established.~\cite{Unify} 
The global rotations of $Q$ with the matrix $\exp(i \hat\chi)$
where $\hat \chi = \sum_{\alpha,n} \chi^\alpha_n I^\alpha_n$ play
the important role.~\cite{Unify,KamenevAndreev} For example,
$\mathcal{F}$-algebra allows us to establish the following
relations
\begin{eqnarray}
\tr I^\alpha_n  t_{ab} e^{i\hat\chi} Q e^{-i\hat\chi} &=& \tr I^\alpha_n  t_{ab} e^{i\chi_0} Q e^{-i\chi_0}
+ 8 i n (\chi_{ab})^\alpha_{-n}\,,\notag\\
\tr \eta e^{i\hat\chi} Q e^{-i\hat\chi} &=& \tr \eta Q +
\sum_{\alpha n;ab } i n
(\chi_{ab})^\alpha_n\tr I^\alpha_n  t_{ab}
Q \notag \\ &-& 4 \sum_{\alpha n;ab}
n^2
(\chi_{ab})^\alpha_n(\chi_{ab})^\alpha_{-n} \label{Falg}
\end{eqnarray}
where $\chi_0 = \sum_{\alpha} \chi_0^\alpha I_0^\alpha$. With the help of Eqs.~\eqref{Falg} one can check that
the relation $\Gamma_s=-z$ guarantees the so-called
$\mathcal{F}$-invariance.~\cite{Unify} It is the invariance of
the action
$\mathcal{S}_\sigma+\mathcal{S}_F$ under the global rotation of the
matrix $Q$ with $\chi_{ab} = \chi \delta_{a0}\delta_{b0}$.

\subsection{Physical observables \label{Sec_Form_PhysObser}}

The most significant physical quantities in the theory containing
information on its low-energy dynamics are physical observables
$\sigma_{xx}^\prime$, $z^\prime$, and $z_{ab}^\prime$ associated
with the mean-field parameters $\sigma_{xx}$, $z$, and $z_{ab}$
of the action~\eqref{Sstart}. The observable $\sigma_{xx}^\prime$
is the total DC conductance as obtained from the linear response
to an electromagnetic field. The observable $z^\prime$ is related
with the specific heat.~\cite{CasDiCas} The observables
$z_{ab}$ determine the static generalized susceptibilities of the 2D electron
system~\cite{CastelChi,Finkelstein} as $\chi_{ab} = 2
z_{ab}^\prime/\pi$. The conductance $\sigma^\prime_{xx}$ can obtained from 
\begin{gather}
\sigma^\prime_{xx}(i\omega_n) = -\frac{\sigma_{xx}}{16 n}\left
\langle\tr[I_{n}^\alpha, Q][I_{-n}^\alpha, Q] \right
\rangle\hspace{1.5cm} \notag 
\\
+ \frac{\sigma_{xx}^2}{64 \mathbb{D} n } \int
d\bm{r}^\prime\langle\langle \tr I_n^\alpha
Q(\bm{r})\nabla Q(\bm{r}) \tr I_{-n}^\alpha
Q(\bm{r}^\prime)\nabla Q(\bm{r}^\prime)\rangle \rangle
\label{SigmaODef}
\end{gather}
after the analytic continuation to the real frequencies: $i\omega_n \to
\omega + i0^+$ at $\omega \to 0$. The expectation values are defined with respect
to the theory~\eqref{Sstart} and $\mathbb{D}=2$ stands for
the spatial dimension. 
The physical observable $z^\prime$ can be extracted from the derivative of the
thermodynamic potential $\Omega$ per the unit volume with respect
to temperature,~\cite{Unify}
\begin{equation}
z^\prime = \frac{1}{2\pi \tr \eta \Lambda}\frac{\partial}{\partial
T}\frac{\Omega}{T}.\label{Defz}
\end{equation}
The observables $z_{ab}^\prime$ are given as
\begin{equation}
z^\prime_{ab} = \frac{\pi}{2 N_r} \frac{\partial^2
\Omega}{\partial \Delta_{ab}^2}\Biggr |_{\Delta_{ab}=0} . \label{Defzsv}
\end{equation}
It is worth mentioning that, alternatively, the
observable parameters $\sigma^\prime_{xx}$, $z_{ab}^\prime$ and
$z^\prime$ can be found from the
background field procedure.

%%%%%%%%%%%%%%%%%%%%%%%%%%%%%%%%%%%%%%%

\section{One-loop renormalization\label{Sec_RG}}

\subsection{Perturbative expansions \label{Sec_RG_Pert}}

To define the theory for the perturbative expansions we use the
``square-root'' parameterization:
\begin{gather}
Q = W+\Lambda\sqrt{1-W^2},\qquad W =
  \begin{pmatrix}
    0 & w \\
    w^\dag & 0
  \end{pmatrix} .\label{Qexp}
\end{gather}
The action~\eqref{Sstart} can be written as the infinite series in
the independent fields $w$ and $w^\dag$. 
At short length scales $L\ll\sqrt{\sigma_{xx}/(z_{ab}\Delta_{ab})}$ which we are interested in, the symmetry breaking term $\mathcal{S}_{SB}$ can be omitted. Then
the propagators for fields $w$ and $w^\dag$ can be written in the following form
%\begin{widetext}
%
\begin{gather}
\langle
[w_{ab}(\bm{q})]_{n_1n_2}^{\alpha_1\alpha_2}
[w_{cd}^\dag(-\bm{q})]_{n_4
n_3}^{\alpha_4\alpha_3}
\rangle
\!\! =\!\! \frac{4}{\sigma_{xx}}
%\,\delta_{ab;cd}
%\delta^{\alpha_1\alpha_3}\delta^{\alpha_2\alpha_4}
%\delta_{n_{12},n_{34}} 
\notag \\
\times D_q(\omega_{12})  \Bigl [
\delta_{n_1n_3}
-\frac{32 \pi  T \Gamma_{ab}}{\sigma_{xx}}
\delta^{\alpha_1\alpha_2} D^{(ab)}_q(\omega_{12}) \Bigr ] 
\notag
\\
\times
\delta_{ab;cd}
\delta^{\alpha_1\alpha_3}\delta^{\alpha_2\alpha_4}
\delta_{n_{12},n_{34}}
,\label{Prop}
\end{gather}
%
%\end{widetext}
where $\omega_{12}=\varepsilon_{n_1}-\varepsilon_{n_2} = 2\pi Tn_{12} = 2\pi T(n_1-n_2)$ and
\begin{equation}
\begin{split}
D_q^{-1}(\omega_n) = q^2+\frac{16 z \omega_n}{\sigma_{xx}} \, , \hspace{1cm}{}\, 
\\ [D^{(ab)}_q(\omega_n)]^{-1} = q^2 +\frac{16
(z+\Gamma_{ab})\omega_n}{\sigma_{xx}} \, . 
\end{split}
\end{equation}
We use the convention that
the Matsubara frequency indices with odd subscripts $n_1, n_3, \dots$ run over
non-negative integers whereas those with even subscripts $n_2,
n_4, \dots$ run over negative integers.

\subsection{Relation of $z_{ab}$ with $z$ and $\Gamma_{ab}$\label{Sec_RG_Rel}}

The dynamical susceptibility $\chi_{ab}(\omega,\bm{q})$ which describes the 
linear response of the system to time-dependent symmetry breaking amplitude $\Delta_{ab}$  
can be obtained from~\cite{Finkelstein}
\begin{equation}
\chi_{ab}(i\omega_n,\bm{q}) =\frac{2 z_{ab}}{\pi} - T z_{ab}^2 \langle \tr
I^\alpha_n t_{ab} Q(\bm{q}) \tr I^\alpha_{-n} t_{ab}
Q(-\bm{q})\rangle \label{SS1}
\end{equation}
by the analytic continuation to the real frequencies: $i\omega_n \to
\omega + i0^+$. In the tree level
approximation Eq.~\eqref{SS1} yields
\begin{equation}
\chi_{ab}(i\omega_n,\bm{q}) = \frac{2 z_{ab}}{\pi} \left (1 -
\frac{16 z_{ab} \omega_n}{\sigma_{xx}} D^{ab}_q(\omega_n)\right ).\label{ChisTreeLevel}
\end{equation}
The action $\mathcal{S}_\sigma+\mathcal{S}_F$ is invariant under the global rotations $Q \to u Q u^{-1}$ with 
$u = \sum_{a=0}^1\sum_{b=0}^3 u_{ab} t_{ab}$. 
This implies that the quantities corresponding to operators $m^{0b}$ and $m^{1b}$ conserve, 
i.e., $\chi_{0b}(\omega,\bm{q}=0)=\chi_{1b}(\omega,\bm{q}=0)=0$. 
In order to be consistent with this physical requirement, the relations
\begin{gather}
z_{ab} = z+\Gamma_t = z(1+\gamma_t), \qquad  a=0,1,\, b =1,2,3,\notag \\
z_{10}= z+\tilde \Gamma_s= z(1+\tilde \gamma_s) .
\label{zsR1}
\end{gather}
should hold. Therefore, renormalization of the interaction amplitudes $\tilde\Gamma_s$ and $\Gamma_t$ 
can be easily found from, e.g., renormalized quantities $z_{01}^\prime$ and $z_{10}^\prime$. 
However, it is not the case for the interaction amplitude $\Gamma_v=z\gamma_v$. 
There is no simple relation between $\Gamma_v$ and  
\begin{equation}
z_v=z_{2b} =z_{3b},\,\qquad b=0,\dots, 3 . \label{zsR2}
\end{equation}
Therefore, the physical observables $\sigma_{xx}^\prime$, $z^\prime$, $\tilde\gamma_s^\prime$, $\gamma_t^\prime$, 
$\gamma_v^\prime$ and $z_v^\prime$ completely determines the renormalization of the
theory~\eqref{Sstart} at short length scales $L \ll \sqrt{\sigma_{xx}/z_{ab}\Delta_{ab}}$.

\subsection{One-loop results\label{Sec_RG_One}}

Evaluation of the conductance according to Eq.~\eqref{SigmaODef} in the one-loop
approximation yields
\begin{gather}
\sigma^\prime_{xx}(i\omega_n) = \sigma_{xx}
-\frac{128 \pi T}{\omega_n \sigma_{xx}  \mathbb{D}}\int \frac{d^\mathbb{D} \bm{p}}{(2\pi)^\mathbb{D}}\, p^2 \sum_{ab}  \Gamma_{ab} \notag 
\sum_{\omega_m>0} \hspace{0.5cm}{}\,
\\
\times
\min\{\omega_m,\omega_n\} D_p(\omega_m+\omega_n) 
D_p(\omega_m) D^{(ab)}_p(\omega_m) .
\label{SR}
\end{gather}
Performing the analytic continuation to the real frequencies,
$i\omega_n\to\omega+i0^+$, one obtains the DC
conductance in the one-loop approximation:
\begin{gather}
\sigma^\prime_{xx} = \sigma_{xx} 
+\frac{32}{\sigma_{xx} \mathbb{D}} \Imag  \int \frac{d^\mathbb{D} \bm{p}}{(2\pi)^\mathbb{D}}\, p^2 \sum_{ab}  \Gamma_{ab} \notag 
\int d\Omega \notag \\
\times \frac{\partial}{\partial\Omega} \Bigl ( \Omega \coth \frac{\Omega}{2T} \Bigr )
[D^R_p(\Omega)]^2 
D^{(ab),R}_p(\Omega) .
\label{sigma2}
\end{gather}
Here $D_p^R(\Omega)$ and $D^{(ab),R}_p(\Omega)$ are retarded propagators 
corresponding to $D_p(\omega_n)$ and $D_p^{(ab)}(\omega_n)$, respectively:
\begin{equation}
\begin{split}
[D_p^R(\Omega)]^{-1} = p^2- (16 z/\sigma_{xx}) \, i\Omega \, , \hspace{1cm}{}\, 
\\ [D^{(ab),R}_p(\Omega)]^{-1} = p^2 -(16
(z+\Gamma_{ab})/\sigma_{xx})\, i \Omega  \, . 
\end{split} \label{CondFinDD}
\end{equation}
We mention that the result~\eqref{sigma2} can be also obtained with the help of the background field
procedure~\cite{Amit} applied to the action~\eqref{SstartSigma}-\eqref{SstartF}.

In order to  compute $z^\prime$, we have to
evaluate the thermodynamic potential $\Omega$. In the one-loop approximation we find
\begin{gather}
T^2\frac{\partial\Omega/T}{\partial T}=  8 N_r T\sum_{\omega_n>0}
\omega_n
 \Bigl [z+\frac{2}{\sigma_{xx}} \sum_{ab} \int  \frac{d^\mathbb{D} \bm{p}}{(2\pi)^\mathbb{D}} \hspace{1cm}\,{}\notag \\
 \times \Bigl [ (z+\Gamma_{ab}) D_p^{(ab)}(\omega_n)-z D_p(\omega_n)
\Bigr ]\Bigr
].\label{z1}
\end{gather}
Following definition~\eqref{Defz}, we obtain from Eq.~\eqref{z1}
\begin{gather}
z^\prime =z +\frac{2}{\sigma_{xx}}\sum_{ab} \Gamma_{ab} \int  \frac{d^\mathbb{D} \bm{p}}{(2\pi)^\mathbb{D}}
D_p(0)  . \label{z2}
\end{gather}
Next, we evaluate in the one-loop approximation the generalized susceptibility $\chi_{ab}(i\omega_n,\bm{q})$  at
$q=0$ and $\omega_n\to 0$. Then, according to Eq.~\eqref{Defzsv}, we find 
\begin{gather}
z^\prime_{ab} = z_{ab}  + \frac{ 32 \pi z_{ab}^2}{\sigma_{xx}^2} \sum_{cd;ef} \left [  \mathcal{C}_{cd;ef}^{ab} \right ]^2 
 \int   \frac{d^\mathbb{D} \bm{p}}{(2\pi)^\mathbb{D}} 
T \sum_{\omega_m>0} \notag \\
\times \Bigl [ 
D^{(ef)}_p(\omega_m) D_p^{(cd)}(\omega_m) - D_p^2(\omega_m) \Bigr ] , \label{zabR}
\end{gather}
where $ \mathcal{C}_{cd;ef}^{ab}$ denotes the structural constants of $SU(4)$:  $[t_{cd}, t_{ef}] = \sum_{ab} \mathcal{C}_{cd;ef}^{ab} t_{ab}$.
%\begin{gather}
%z^\prime_{ab} = z_{ab}  - \frac{ 4\pi z_{ab}^2}{\sigma_{xx}^2} \sum_{cd;ef}    \Sp [t_{ab} t_{cd} t_{ef}] \hspace{3cm}\,{}
%\notag \\
%\times \Bigl \{  \Sp [t_{ab} t_{ef} t_{cd}]  -  \Sp [t_{ab} t_{cd} t_{ef}]  \Bigr \} \int   \frac{d^d \bm{p}}{(2\pi)^d} 
%T \sum_{\omega_m>0} \notag \\
%\times \Bigl [ 
%D^{(ef)}_p(\omega_m) D_p^{(cd)}(\omega_m) - D_p^2(\omega_m) \Bigr ] . \label{zabR}
%\end{gather}
Applying Eq.~\eqref{zabR} for $(ab) = (10)$ and $(ab)=(01)$, and by virtue of relations~\eqref{zsR1}
we obtain
\begin{gather}
z^\prime+\tilde{\Gamma}_s^\prime = z+\tilde{\Gamma}_s - \frac{2^{10}\pi (z+\tilde{\Gamma}_s)^2}{\sigma_{xx}^2}
\int   \frac{d^\mathbb{D} \bm{p}}{(2\pi)^\mathbb{D}} 
T\sum_{\omega_m>0} \notag \\
\times \Bigl \{ \bigl [D^{(20)}_p(\omega_m) \bigr ]^2- D_p^2(\omega_m) \Bigr \} ,
\label{zst2}
\\
z^\prime+\Gamma_t^\prime = z+\Gamma_t - \frac{2^9 \pi (z+\Gamma_t)^2}{\sigma_{xx}^2}
\int   \frac{d^\mathbb{D} \bm{p}}{(2\pi)^\mathbb{D}} 
T \sum_{\omega_m>0} \notag \\
\times \Bigl \{ \bigl [D^{(01)}_p(\omega_m) \bigr ]^2 +  \bigl [D^{(20)}_p(\omega_m) \bigr ]^2 - 2 D_p^2(\omega_m) \Bigr \} .
\label{zt2}
\end{gather}

In order to find renormalization of $\Gamma_v$, one cannot use the static generalized susceptibility since there exists no simple relation between $z_v$ and $\Gamma_v$. We use the the background-field renormalization procedure (see details in Appendix~\ref{BGR}) and find
\begin{gather}
\Gamma_{ab}^\prime = \Gamma_{ab} - \frac{1}{8 \sigma_{xx}} \int  \frac{d^\mathbb{D} \bm{p}}{(2\pi)^\mathbb{D}} D_p(0) \sum_{cd;ef}
\Gamma_{cd} \bigl [  \Sp (t_{cd} t_{ef} t_{ab}) \bigr ]^2 \notag \\
- \frac{32 \pi T}{\sigma_{xx}^2} \sum_{\omega_m>0} \int  \frac{d^\mathbb{D} \bm{p}}{(2\pi)^\mathbb{D}}
\sum_{cd;ef} \left [  \mathcal{C}_{cd;ef}^{ab} \right ]^2 \Bigl \{\Gamma_{ab}^2 D_p^2(\omega_m) \notag \\
-\Bigl [ \Gamma_{cd}\Gamma_{ef}+\Gamma_{ab}^2-2\Gamma_{ab}\Gamma_{cd}\Bigl ]
 D_p^{(cd)}(\omega_m)D_p^{(ef)}(\omega_m)  \Bigr \} .\notag \\ \label{RenGam1}
\end{gather}
Here symbol $\Sp$ denotes trace over spin and isospin indices. Using Eq.~\eqref{RenGam1} for $(ab) = (02)$ , we obtain
\begin{eqnarray}
\Gamma^\prime_v = \Gamma_v &-& \frac{2(\Gamma_s-\tilde{\Gamma}_s)}{\sigma_{xx}} \int  \frac{d^\mathbb{D} \bm{p}}{(2\pi)^\mathbb{D}} D_p(0) \notag \\
&+& \frac{2^{10} \pi \Gamma_v^2}{\sigma_{xx}} T \sum_{\omega_m>0} \int  \frac{d^\mathbb{D} \bm{p}}{(2\pi)^\mathbb{D}} D_p^2(\omega_m) .
\label{zv2}
\end{eqnarray}
It is worthwhile to mention that the results~\eqref{z2},
\eqref{zst2} and \eqref{zt2} can be also derived  from Eq.~\eqref{RenGam1}. Equations~\eqref{sigma2}~\eqref{z2}, \eqref{zst2}, \eqref{zt2} and \eqref{zv2} allow us to extract one-loop renormalization of conductance $\sigma_{xx}$, 
parameter $z$ and interaction amplitudes $\tilde{\Gamma}_s$, $\Gamma_t$ and $\Gamma_v$.  
%
%
%
%%%%%%%%%%%%%%%%%%
\subsection{Renormalization group equations\label{Sec:RG:RGE}}

Applying the minimal subtraction scheme (see, e.g., Ref. \onlinecite{Amit}) to
Eqs.~\eqref{sigma2}, \eqref{z2}, \eqref{zt2}, \eqref{zst2} and \eqref{zv2}, we derive the following one-loop results
for the renormalization group (RG) equations which determine the $T=0$ behavior of the
physical observables with changing the length scale $L$ in $\mathbb{D}=2$ dimensions:
\begin{gather}
\frac{d \sigma_{xx}}{d\xi} =- \frac{2}{\pi}\bigl [1+ f(\tilde{\gamma}_{s})+6 f(\gamma_t)+
8 f(\gamma_{v})  \bigr ] ,\label{RG1_1}\\
\frac{d\tilde{\gamma}_s}{d\xi} =
\frac{1+\tilde{\gamma}_s}{\pi\sigma_{xx}}\Bigl [ 1-6\gamma_t-\tilde{\gamma}_s +8\gamma_v+16 \gamma_v \frac{\tilde{\gamma}_s-\gamma_v}{1+\gamma_v} \Bigr ] ,\label{RG1_2}\\
\frac{d{\gamma}_t}{d\xi} =
\frac{1+\gamma_t}{\pi\sigma_{xx}}\Bigl [1-\tilde{\gamma}_s +2\gamma_t +8 \gamma_v \frac{\gamma_t-\gamma_v}{1+\gamma_v}  \Bigr ] ,
\label{RG1_3}\\
\frac{d\gamma_v}{d\xi} =
\frac{1}{\pi\sigma_{xx}} \Bigl [ 1+\tilde\gamma_s+\gamma_v-\gamma_v(6\gamma_t+\tilde{\gamma}_s) +8 \gamma_v^2\Bigr  ] ,
\label{RG1_4}\\
\frac{d\ln z}{d\xi} = \frac{1}{\pi\sigma_{xx}} \Bigl
[\tilde{\gamma}_s+6\gamma_t+8\gamma_v-1 \Bigr ].\label{RG1_5}
\end{gather}
Here $f(x) = 1- (1+x^{-1})\ln(1+x)$, $\xi=\ln L/l$ and we omit primes for a brevity. 
Equations~\eqref{RG1_1}-\eqref{RG1_4} constitute one of the main results of the present paper and describe the system at the
length scales $L\ll \sqrt{\sigma_{xx}/(z_{ab}\Delta_{ab})}$.

It is worthwhile to mention that the right hand side of Eqs.~\eqref{RG1_2} and \eqref{RG1_3} is not polynomial in the interaction amplitude $\gamma_v$. To the best of our knowledge, the one-loop RG equations for interaction amplitudes are quadratic polynomials in all cases studied previously.~\cite{Finkelstein, KirkpatricBelitz, BurmistrovChtchelkatchev,Punnoose2, Punnoose1} 
This fact is deeply related with  invariance of the action $\mathcal{S}_\sigma+\mathcal{S}_F$ under the global rotation of the
matrix $Q$ with the matrix $\exp(i\hat\chi)$ (see Sec.~\ref{Sec:Formalism:Finv}). As it follows from Eqs.~\eqref{Falg}, 
$\mathcal{S}_\sigma+\mathcal{S}_F$ is invariant under such global rotation with $\chi_{ab} = \chi \delta_{ac}\delta_{bd}$ where $c=0,1$ and $d=1,2$ or $3$ provided $\gamma_t=-1$. The same holds for the global rotation with $\chi_{ab} = \chi \delta_{a1}\delta_{b0}$ if $\tilde{\gamma}_s=-1$.
This invariance guarantees that $\gamma_t=-1$ and  $\tilde{\gamma}_s=-1$ are fixed points of the RG equations. Therefore, the latter have to be well-defined at $\gamma_t=-1$ and $\tilde{\gamma}_s=-1$. However, for $\gamma_v=-1$ the action $\mathcal{S}_\sigma+\mathcal{S}_F$ is not invariant under the global rotation of the matrix $Q$ with $\chi_{ab} = \chi \delta_{ac}\delta_{bd}$ with $c=1,2$ and $d=0,1,2$ or $3$.  
It is exactly this noninvariance that allows appearance of factors $1/(1+\gamma_v)$ (diverging at $\gamma_v=-1$) in Eqs.~\eqref{RG1_2} and \eqref{RG1_3}.

The renormalization group equations~\eqref{RG1_1}-\eqref{RG1_4} possess a rich four-dimensional ($\sigma_{xx}, \tilde\gamma_s, \gamma_t, \gamma_v$) flow diagram. First of all, there is the two-dimensional surface $\gamma_t=\gamma_v=\tilde\gamma_s$ which is conserved under RG flow. It corresponds to the case of coinciding quantum wells ($d=0$). In this case, the RG equations ~\eqref{RG1_1}-\eqref{RG1_5} are completely equivalent to ones for the two-valley electron liquid. However, this two-dimensional surface is unstable: a small initial mismatch (e.g., due to finite $d$) in the condition $\gamma_t=\gamma_v=\tilde\gamma_s$ increases during RG flow. Secondly, the RG flow conserves the two-dimensional surface $\gamma_v=0$, $\tilde\gamma_s=-1$ which is stable. It describes the limit of two separate quantum wells ($d=\infty$). In addition, there are some interesting features of RG flow. For example, there is a two-dimensional surface $\gamma_t=\tilde\gamma_s=-1$ which is conserved by RG flow. 
There is an accidental fixed line  $\tilde\gamma_s = -1$, $\gamma_v=-1/2$, $\gamma_t=-1/3$. 
However, these features are not accessible in the double quantum well structure. 

\begin{figure}[t]
\centerline{\includegraphics[width=80mm]{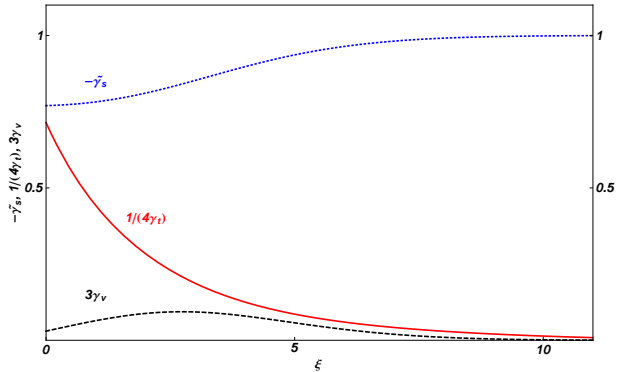}}
\caption{Dependence of the parameters $\gamma_t$, $\gamma_v$, $\tilde\gamma_s$ on $\xi$. Initial values are
$\bar\gamma_t=0.35$, $\bar \gamma_v=0.01$, $\bar{\tilde\gamma}_s
=-0.77$, and $\bar\sigma_{xx}=6$.}
\label{Gammas_Figure}
%\vspace{-0.5cm}
\end{figure}

Indeed, the initial values of the parameters $\gamma_t$, $\gamma_v$ and $\tilde\gamma_s$ satisfy
\begin{equation}
\bar\gamma_t \geqslant \bar \gamma_v\geqslant 0, \qquad \bar\gamma_t \geqslant \bar{\tilde\gamma}_s . \label{CondRGIn}
\end{equation} 
Then, using Eq.~\eqref{RG1_2}-\eqref{RG1_4} one can prove that under RG flow i) the conditions $\gamma_t \geqslant \gamma_v\geqslant 0$ and $\gamma_t \geqslant \tilde\gamma_s$ hold; ii) $\gamma_t$ always increases. Starting from initial values of the parameters $\gamma_t$, $\gamma_v$ and $\tilde\gamma_s$ satisfying Eq.~\eqref{CondRGIn} the RG flow develops in such a way that $\gamma_v$ vanishes, $\tilde\gamma_s$ tends to $-1$ and $\gamma_t$ increases towards infinity as shown in Fig.~\ref{Gammas_Figure}. 

\begin{figure}[t]
\centerline{\includegraphics[width=80mm]{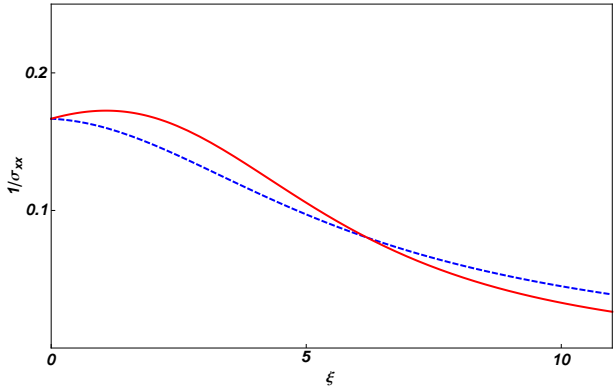}}
\caption{Dependence of the inverse conductance $1/\sigma_{xx}$ on $\xi$. Solid (red) curve is plotted with initial values 
$\bar\gamma_t=0.35$, $\bar \gamma_v=0.01$, $\bar{\tilde\gamma}_s
=-0.77$, and $\bar\sigma_{xx}=6$. Dashed (blue) curve corresponds to $\bar\gamma_t=0.198$, $\bar \gamma_v=0.135$, $\bar{\tilde\gamma}_s
=0.57$, and $\bar\sigma_{xx}=6$.}
\label{Resistance_Figure}
%\vspace{-0.5cm}
\end{figure}

The conductance $\sigma_{xx}$ demonstrates metallic behavior as in the case of two-valley electron liquid. It increases at large length scales. Depending on the sign of the parameter $K_{ee} =  1+f(\bar{\tilde \gamma}_s)+6f(\bar\gamma_t) +8 f(\bar \gamma_v)$,
the conductance can develop both monotonic ($K_{ee}<0$) and non-monotonic behavior ($K_{ee}>0$) (see Fig.~\ref{Resistance_Figure}).  The phase diagram for the parameter $K_{ee}$ is shown in Fig.~\ref{ResisPM_Figure}. At $\varkappa/k_F\lesssim 0.4 $ the parameter $K_{ee}$ is positive for all values of $\varkappa d$. 
With increasing $\varkappa/k_F$ a domain of negative values of $K_{ee}$ develops at small values of $\varkappa d$. 

%%%%%%%%%%%%%%%%%%%%%%%%%%%%%%%%%%%%%%%%%%%%%%%%%%%%%%%%%%%%%%%%%%%%%%%%%%%%%
%%%%%%%%%%%%%%%%%%%%%%%%%%%%%%%%%%%%%%%%%%%%%%%%%%%%%%%%%%%%%%%%%%%%%%%%%%%

 The conductance $\sigma^\prime_{xx}$ defined in Eq.~\eqref{SigmaODef} and renormalized in accordance with 
Eq.~\eqref{RG1_1} is the total conductance of double quantum well structure. In general, one can write $\sigma_{xx}^\prime = \sigma_{11}^\prime+\sigma_{22}^\prime +\sigma_{12}^\prime+\sigma_{21}^\prime$, where $\sigma_{11}^\prime$ and $\sigma_{22}^\prime$ 
are the intrawell conductances of left and right quantum wells respectively, and $\sigma_{12}^\prime$ and $\sigma_{21}^\prime$ denote the transconductances responsible for a drag effect. At the balance, symmetry yields that $\sigma_{11}^\prime=\sigma_{22}^\prime$ and $\sigma_{12}^\prime=\sigma_{21}^\prime$. 

Although in experiments of Refs.~[\onlinecite{MinkovGermanenko1,MinkovGermanenko2}] only the total conductivity  $\sigma_{xx}^\prime$ has been measured, such double quantum well heterostructures with correlated disorder at the balance allow for experimental study of transconductance contrary to the two-valley electron system in Si-MOSFET. It was shown~\cite{DragIgor} that in the presence of electron-electron interaction one-loop contribution  in the particle-hole channel (only ``diffusons'') to the DC transconductance  $\sigma_{12}^\prime$ vanishes. As a result, the one-loop contribution to the DC transconductance is entirely determined by the particle-particle channel (``Cooperons''). 
However, in Ref.~[\onlinecite{DragIgor}] only the interwell interactions ($U_{12}^{\rm scr}$) were taken into account. As shown in Appendix~\ref{TC}, an accurate treatment of both interwell ($U_{12}^{\rm scr}$) and intrawell ($U_{11}^{\rm scr}$) interactions (i.e. taking into account all interaction couplings $\Gamma_s$, $\tilde\Gamma_s$, $\Gamma_t$, and $\Gamma_v$) does not change the conclusion of Ref.~[\onlinecite{DragIgor}]: the particle-hole  (``diffuson'') contribution to the DC transconductance $\sigma_{12}^\prime$ vanishes in the one-loop approximation.

%%%%%%%%%%%%%%%%%%%%%%%%%%%%%%%%%%%%%%%

\section{Dephasing rate\label{Sec_Deph}}

The presence of the right well changes the properties of electrons in the left well. One of the important quantities characterizing interacting electrons in a random potential is the dephasing rate. Its dependence on temperature determines the behavior of the weak-localization correction to the conductance. In this section, we investigate how the presence of the right well changes the dephasing rate of electrons in the left well compared to the case when the right well is empty.

\subsection{Contribution from the interaction in the singlet channel} 

We start from the case of the interaction in the singlet channel only. According to Eq.~\eqref{ScrU11}, electrons in the right well screen interaction between electrons in the left well and vice versa. The dephasing rate of electrons in the left well due to the presence of electrons in the right well 
can be found from the following expression which generalizes standard on:~\cite{Aronov-Altshuler,Schmid}
\begin{equation}
\frac{1}{\tau_\varphi} = -\int_{\tau_\varphi^{-1}} \frac{d\omega}{\pi} \int \frac{d^2 q}{(2\pi)^2} \frac{\Im U_{11}^{\rm scr}(\bm{q},\omega)}{\sinh{(\omega/T)}}
%\Re 
 \frac{D_1 q^2}{D^2_1q^4+\omega^2}
%  \hspace{1cm}\,{}\notag \\
% \times  \Re\left (  \frac{1}{D_1q^2-i\omega}\right ) 
. \label{Int}
\end{equation}
Expression for the dephasing rate of electrons in the right well can be obtained from Eq.~\eqref{Int} by substitution of $U_{22}$ and $D_2$ for $U_{11}$ and $D_1$, respectively. At the balance which we are interested in, the dephasing rates in the left and right wells are the same.  Under the following assumption $d,\varkappa^{-1}\ll L_T=\sqrt{D/T}$, we find
\begin{equation}
\frac{1}{\tau_\varphi} = \mathcal{A}_s \frac{T}{8\pi \nu D} \ln T \tau_\varphi 
\label{TauPhi1}
\end{equation}
where $A_s$ is the function of the parameter $\varkappa d$: 
\begin{equation}
\mathcal{A}_s = \frac{1}{2} \left [ 1+  \frac{(\varkappa d)^2}{(1+\varkappa d)(2+\varkappa d)} \right ] . \label{As}
\end{equation}
We mention that in the absence of electrons in the right well (formally this case corresponds to the limit $d\to \infty$) the dephasing rate is maximal: $\mathcal{A}_s=1$. Equation~\eqref{As} was used for analysis of the experimental data in Ref.~[\onlinecite{MinkovGermanenko1}].

%%%%%%%%%%%%%%%%%%%%%%%%%%%%%%%%%%%%%%%%%%%%%%%%%%%%%%%%%%%%%%%%%%%%%%%%%%%%%
%%%%%%%%%%%%%%%%%%%%%%%%%%%%%%%%%%%%%%%%%%%%%%%%%%%%%%%%%%%%%%%%%%%%%%%%%%%

%%%%%%%%%%%%%%%%%%%%%%%%%%%%%%%%%%%%%%%%%%%%%%%%%%%%%%%%%%%%%%%%%%%%%%%%%%%%%
\begin{figure}[t]
\centerline{\includegraphics[width=75mm]{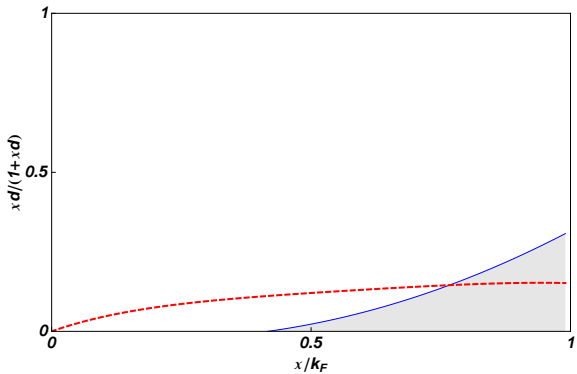}}
\caption{The phase diagram for the parameter $K_{ee}=1+f(\bar{\tilde \gamma}_s)+6f(\bar\gamma_t) +8 f(\bar \gamma_v)$. It vanishes on the solid (blue) line. 
$K_{ee}$ is negative in the filled region below the solid (blue) curve and is positive above. Dashed (red) line indicates $\tilde F_s=0$. }
\label{ResisPM_Figure}
%\vspace{-0.5cm}
\end{figure}
%%%%%%%%%%%%%%%%%%%%%%%%%%%%%%%%%%%%%%%%%%%%%%%%%%%%%%%%%%%%%%%%%%%%%%%%%%%

\subsection{Contribution from the interaction in the multiplet channels}

In general case, one has to take into account contributions to the dephasing rate from the interaction in multiplet channels.~\cite{Narozhny} 
We restrict ourselves to the case of the balance. Generalizing the well-known result~\cite{Aronov-Altshuler} for the single well  we can write the dephasing rate in the left well as 
\begin{gather}
\frac{1}{\tau_\varphi} = -\frac{2}{\sigma_{xx}}\int_{\tau_\varphi^{-1}} \!\!\!\!\!d\omega \int \frac{d^2 q}{(2\pi)^2} \frac{\Re D_q^R(\omega)}{\sinh{(\omega/T)}}
% \Re D_q^R(\omega)  \hspace{1cm}\,{}\notag \\
% \times 
\sum_{ab} \Im\, \mathcal{U}^{(ab)}(q,\omega) \label{Int1}
\end{gather}
where
\begin{equation}
\mathcal{U}^{(ab)}(q,\omega) = \frac{\Gamma_{ab}}{z} D_q^{(ab),R}(\omega) [D_q^R(\omega)]^{-1} .
\end{equation}
Performing integration over momentum and frequency, we find
\begin{equation}
\frac{1}{\tau_\varphi} = \mathcal{A} \frac{T}{2\sigma_{xx}} \ln T \tau_\varphi 
\label{TauPhi2}
\end{equation}
with
\begin{equation}
\mathcal{A} = \frac{1}{2} \left [ 1 + \frac{\tilde{\gamma}_s^2}{2+\tilde{\gamma}_s}+6\frac{\gamma_t^2}{2+\gamma_t} +8 \frac{\gamma_v^2}{2+\gamma_v} \right ] .
\end{equation}
In the absence of interaction in the multiplet channels, i.e., for $F_t=F_v=0$ and $\tilde\gamma_s=-\varkappa d/(1+\varkappa d)$, this result transforms into Eq.~\eqref{TauPhi1}. We mention that the interaction parameters $\tilde \gamma_s$, $\gamma_t$ and $\gamma_v$ as well as conductance $\sigma_{xx}$ should be taken at the length scale $L_T = \sqrt{\sigma_{xx}/z T}$.

It is worthwhile to compare Eq.~\eqref{TauPhi2} with the result for the dephasing rate in the absence of electrons in the right well.~\cite{Aronov-Altshuler}
Taking the limit $d\to \infty$, i.e., setting $\tilde\gamma_s=-1$, $\gamma_v=0$, and $\gamma_t = \gamma_{t,0}$, we obtain
\begin{equation}
\mathcal{A} \,\to\, \mathcal{A}_0=  \left [ 1 + \frac{3\gamma_{t,0}^2}{2+\gamma_{t,0}} \right ]
\end{equation}
where initial value of $\gamma_{t,0}$ is $\bar\gamma_{t,0}= -F_t^0/(1+F_t^0)$.

%%%%%%%%%%%%%%%%%%%%%%%%%%%%%%%%%%%%%%%

\section{Comparison with the experiment\label{Sec_Disc}}

Recently, the interference~\cite{MinkovGermanenko1} and interaction~\cite{MinkovGermanenko2}
corrections to the conductivity of the gated double quantum well Al$_x$Ga$_{1-x}$As/GaAs/Al$_x$Ga$_{1-x}$As heterostructures have been studied.
Two heterostructures, 3243 and 3154, distinguishing by the doping level have been investigated. From analysis of positive magnetoconductivity the dephasing rate has been extracted. By tuning the gate voltage, the electron concentration in the right quantum well were controlled in the experiment. 

We consider two characteristic cases: I) electron concentrations and mobilities ($\mathcal{M}$) of both quantum wells are equal: $n_1=n_2=n$ and $\mathcal{M}_1=\mathcal{M}_2=\mathcal{M}$; II) the left quantum well has electron concentration $n_1=n$ and mobility $\mathcal{M}_1=\mathcal{M}$ whereas the right quantum well has electron concentration $n_2=0$. The electron concentration $n$ has been high such that the conductance $\bar\sigma_{xx}$ was about $80$. Therefore, physics described by RG equations~\eqref{RG1_1}-\eqref{RG1_5} was not observed. The main unexpected findings of Refs.~[\onlinecite{MinkovGermanenko1,MinkovGermanenko2}] were as follows. Dephasing rates (coefficient $\mathcal{A}$) and interaction correction (parameter $K_{ee}$) extracted in cases I) and II) were practically the same. At first glance, it is counterintuitive since there are 15 multiplets in the case I) and only $3$ in the case II). 

After Refs.~[\onlinecite{MinkovGermanenko1,MinkovGermanenko2}] we summarize the experimental values of relevant parameters in 
Table~\ref{Tab1}. The theoretical estimates for the interaction parameters in cases I) and II) are presented in Table~\ref{Tab2}. As one can see, in 
the experimentally studied case of $\varkappa d=3.6$ the interaction parameter $F_v$ is negligible, $F_t$ and $F_t^0$ coincide with each other and $\tilde F_s$ is equal approximately to $\varkappa d$. The comparison between theoretical estimates for $K_{ee}$, $\mathcal{A}$, $K_{ee,0}$ and $\mathcal{A}_0$ with experimental data (whereever it is possible) is summarized in Table~\ref{Tab3}. Our theoretical estimates are in good quantitative agreement with the experimental ones. Our results explain why the interaction corrections and dephasing rates in cases I) and II) were found to be practically the same in the experiments.~\cite{MinkovGermanenko1,MinkovGermanenko2} Since the parameter $K_{ee,0}$ is positive for $\varkappa/k_F\lesssim 1$ a drastic effect in the interaction correction could be seen by tuning the gate voltage from case I) to case II) in double quantum well structures with $\varkappa d \lesssim 1$ for which one can expect $K_{ee} < 0$ (see Fig.~\ref{ResisPM_Figure}).

%%%%%%%%%%%%%%%%%%%%%%%%
\begin{table}[t]
\caption{Parameters for samples studied in Refs.~[\onlinecite{MinkovGermanenko1,MinkovGermanenko2}]}
\begin{tabular}{ccc}
sample &\hspace{0.5cm}  \#3154\hspace{0.5cm} & \#3243 \\
\hline  & & \\
$n$, $10^{11}$ cm$^{-2}$ & $4.5$ & $7.5$ \\ 
$k_F$, $10^{6}$ cm$^{-1}$  & $1.7$ & $2.2$\\
$\varkappa$, $10^{6}$ cm$^{-1}$ & $2$ & $2$ \\
$d$, $10^{-6}$ cm & $1.8$ & $1.8$ \\
$\varkappa d$ & $3.6$ & $3.6$ \\
$k_F d$ & $3.06$ & $3.95$ \\
$\varkappa/k_F$ & $1.18$ & $0.91$\\
\hline && \\
\end{tabular}\label{Tab1}
\end{table}
%%%%%%%%%%%%%%%%%%%%%%%%%

%%%%%%%%%%%%%%%%%%%%%%%%%
\begin{table}[t]
\caption{Theoretical estimates of interaction parameters.}
\begin{tabular}{ccc}
sample &\hspace{0.5cm}  \#3154\hspace{0.5cm} & \#3243 \\
\hline  & & \\
$\tilde F_s$ & $3.34$ & $3.37$ \\
$F_t$ & $-0.26$ & $-0.23$ \\
$F_v$ & $-0.009$ & $-0.007$ \\
$\bar{\tilde \gamma}_s$ & $-0.77$ & $-0.77$ \\
$\bar{\gamma}_t$ & $0.35$ & $0.30$ \\
$\bar{\gamma}_v$ & $0.009$ & $0.007$ \\
\hline 
$F_t^0$ &$-0.26$ &$-0.23$\\
$\bar{\gamma}_{t,0}$ &$0.35$ & $0.30$\end{tabular}\label{Tab2}
\end{table}
%%%%%%%%%%%%%%%%%%%%%%%%

%%%%%%%%%%%%%%%%%%%%%%%%%
\begin{table}[b]
\caption{Comparison of theoretical estimates and experimental findings [$K_{ee,0}=1+3f(\bar{\gamma}_{t,0})$, $K_{ee}=1+f(\bar{\tilde\gamma}_s)+6f(\bar\gamma_t)+8f(\bar\gamma_v)$].}
\begin{tabular}{c|cc|cc}
& \hspace{.9cm}Theory\hspace{-.9cm} & && \hspace{-1.9cm}Experiment \hspace{1.9cm}\\
 &\hspace{0.5cm}  \#3154\hspace{0.5cm} & \#3243 &\hspace{0.5cm}  \#3154\hspace{0.5cm} & \#3243 \\
\hline  & & &&\\
$K_{ee}$ & $0.59$ & $0.72$ &$0.50\pm 0.05$& $0.57\pm 0.05$ \\
$K_{ee,0}$ &$0.52$& $0.59$ &$0.53\pm 0.05$& $0.60 \pm 0.05$ \\
$\mathcal{A}$ & $0.89$ & $0.86$ && \\
$\mathcal{A}_0$ &$1.15$ & $1.12$ && \\ 
$\mathcal{A}/\mathcal{A}_0$ & 0.77 & 0.77 & $1.00\pm 0.05$ & $1.00\pm 0.05$
\end{tabular}\label{Tab3}
\end{table}

 As mentioned in the Introduction, our theory is valid at temperatures $T\gg \Delta_{SAS}, \Delta_s, 1/\tau_{+-}$. 
In the experiments of Refs.~[\onlinecite{MinkovGermanenko1,MinkovGermanenko2}] the Zeeman splitting (at relevant magnetic field which was used in order to extract interaction correction) and $\Delta_{SAS}$ were estimated  as $\Delta_s \lesssim 0.2 K$ and $\Delta_{SAS} \lesssim 1K$. A small asymmetry in the impurity distribution along $z$ axis presented in the double quantum well heterostructures used in Refs.~[\onlinecite{MinkovGermanenko1,MinkovGermanenko2}] leads to appearance of scattering rate between symmetric and antisymmetric states.
The corresponding scattering rate ($1/\tau_{+-}$) can be estimated from temperature and magnetic field dependence of weak-localization (interference) correction to conductivity. 

As known,~\cite{AALK81,AA81,Aronov-Altshuler} in the absence of scattering between symmetric and antisymmetric states neither $\Delta_s$ nor $\Delta_{SAS}$ does not influence the weak-localization contribution. In the absence of magnetic field, the weak localization correction to the conductance in both asymptotic cases $\Delta_{SAS}\ll 1/\tau_{+-}$ and $\Delta_{SAS}\gg1/\tau_{+-}$
can be written as 
\begin{equation}
\delta\sigma_{xx}^{WL} = \frac{1}{\pi} \ln \Bigl [ \frac{\tau_{\rm tr}^2}{\tau_\varphi}\Bigl (\frac{1}{\tau_\varphi}+\frac{1}{\tau_{12}}\Bigr )\Bigr ] \label{WL1}
\end{equation}
where $1/\tau_{12} \sim \min \{\Delta_{SAS}^2\tau_{+-},1/\tau_{+-}\}$. The temperature dependence of the weak-localization correction~\eqref{WL1} smoothly interpolates between the result known for a two-valley system at high temperatures ($1/\tau_\varphi\gg 1/\tau_{12}$) and the single-valley result at low temperatures ($1/\tau_\varphi\ll 1/\tau_{12}$). In experiments~[\onlinecite{MinkovGermanenko1}] the characteristic time $\tau_{12}$ was estimated from the suppression of weak-localization correction due to perpendicular magnetic field as $1/\tau_{12} \lesssim 0.1 K$. Together with the estimate $\Delta_{SAS} \lesssim 1K$ it implies that $1/\tau_{+-} \sim 1/\tau_{12}\lesssim 0.1K$. Therefore our theory is applicable at temperatures $T\gtrsim 1K$. It is this temperature range that was studied  experimentally in Refs.~[\onlinecite{MinkovGermanenko1,MinkovGermanenko2}].

%%%%%%%%%%%%%%%%%%%%%%%%%%

\section{Conclusions\label{Sec_Conc}}

To summarize, we have developed the theory of the disordered electron liquid in a double well quantum heterostructure with equal electron concentrations. 
We have identified all relevant interaction parameters and found their dependence on the distance between quantum wells. To describe the system at low temperatures, we have derived the interacting non-linear sigma model and studied it renormalization in the one-loop approximation. We have obtained the renormalization group equations describing the length scale dependence of the conductance and interaction parameters. We have found that upon the renormalization the system flows towards the fixed point corresponding to two separate quatum wells. The RG equations predict the metallic behavior of the conductance.   
We have evaluated the dephasing rate of electrons due to the presence of electron-electron interaction. This expression takes into account screening of electron-electron interaction within one quantum well by electrons from the other quantum well. 

We did not consider contributions to the one-loop RG equations from the particle-particle (``Cooper")
channel. The interaction effects related to the Cooper channel are governed by the corresponding interaction amplitude which is always small for 2D electron systems with Coulomb repulsion, so that one can neglect it.~\cite{Finkelstein}
As for the interference contribution to conductance, for $1/\tau_\varphi \gg 1/\tau_{12}$, it can be taken into account by the
substitution of $1+2$ for $1$ in the square brackets of Eq.~\eqref{RG1_1}. 
This does not change qualitative behavior of the interaction amplitudes $\tilde\gamma_s$, $\gamma_t$ and $\gamma_v$ discussed above. However, the interference contribution makes behavior of the conductance always non-monotonous.

We performed detailed comparison between our theory and experimental data.~\cite{MinkovGermanenko1,MinkovGermanenko2} We explained main experimental results and found good quantitative agreement. It would be an experimental challenge to construct the double quantum well heterostructure with $\varkappa d\lesssim 1$. Then, according to our predictions, one can expect a change from non-monotonous to monotonous behavior in conductance in the presence of small perpendicular magnetic field (to suppress interference contribution) when the right well is depopulated by tuning the gate voltage. It would be also interesting to experimentally study the Coulomb drag effect in such heterostructures with correlated disorder.

Finally it would be worthwhile to extend our analysis to temperatures less than the symmetry breaking energy scales $\Delta_{SAS}$, $\Delta_s$ and $1/\tau_{+-}$. At such low temperatures one may expect different behavior of transport in double quantum well structures as compared to two-valley electron systems studied recently.~\cite{BurmistrovChtchelkatchev,Punnoose1,Punnoose2}

%%%%%%%%%%%%%%%%%%%%%%%%%%%%%%%%%%%%%%%

\begin{acknowledgements}

The authors thank A. Ioselevich and A. Yashenkin for useful discussions, and are grateful to G. Minkov, A. Germanenko and A. Sherstobitov for detailed description of their experimental data prior to publication. The research was funded in part by 
the Council for Grant of the President of Russian Federation (Grant No. MK-125.2009.2), RFBR
(Grant Nos 09-02-12206 and 09-02-00247-a), RAS Programs ``Quantum Physics of Condensed Matter'' and ``Fundamentals of
nanotechnology and nanomaterials'', the Russian
Ministry of Education and Science under contract No. P926, by the Center for Functional Nanostructures of the Deutsche
Forschungsgemeinschaft, by the DFG-RFBR cooperation grant, and by the EUROHORCS/ESF EURYI Awards scheme. 
The work by K.S.T. was supported by Dynasty Foundation.
I.S.B. is grateful to the Institute of Nanotechnology and Institute of Condensed Matter Theory
at Karlsruhe Institute of Technology for hospitality. 

\end{acknowledgements}

%%%%%%%%%%%%%%%%%%%%%%%%%%%%%%%%%%

\appendix

\section{Background field renormalization of the Finkelstein term $\mathcal{S}_F$ \label{BGR}}

In this appendix we present details of the derivation of Eq.~\eqref{RenGam1} with the help of the background field renormalization. 
Let us separate the matrix field $Q$ into the ``fast'' ($Q$) and ``slow'' ($Q_0=T_0^{-1} \Lambda T_0$) modes as
\begin{equation}
Q \to  T^{-1}_{0} Q T_{0} . \label{Qhat}
\end{equation}
The effective action for the $Q_0$ fields is given by
\begin{equation}
\exp \mathcal{S}_{\rm eff}[Q_0] = \int \mathcal{D}[Q] \exp \mathcal{S}[T_0^{-1} Q T_{0}]
\label{SeffA}
\end{equation}
Since we are interesting in the renormalization of the interaction parameters $\Gamma_{ab}$ only, we insert the spatial independent background field $T_{0}$ in the action~\eqref{SstartF}. The result can be written as follows
\begin{gather}
\mathcal{S}_F[T_0^{-1} Q T_{0}] = \mathcal{S}_F[Q_0] +\mathcal{S}_F[Q] + O_t^{(1),1}+O_t^{(1),2}\notag \\
+O_t^{(2),1}+O_t^{(2),2}+Q_\eta ,
\end{gather}
where
\begin{eqnarray}
O_t^{(1),1} &=&- \frac{\pi T}{2} \int d \bm{r}\sum_{\alpha n; a b} \bm{\Gamma}_{ab} \tr
I_n^\alpha t_{ab} \delta Q \tr I_{-n}^\alpha t_{ab} Q_0  , \notag\\
O_t^{(1),2} &=&- \frac{\pi T}{2} \int d \bm{r}\sum_{\alpha n; a b} \bm{\Gamma}_{ab} \tr
I_n^\alpha t_{ab} \delta Q \tr A_{-n;ab}^\alpha \delta Q  , \notag \\
O_t^{(2),1} &=&- \frac{\pi T}{2} \int d \bm{r}\sum_{\alpha n; a b} \bm{\Gamma}_{ab} \tr
I_n^\alpha t_{ab} Q_0 \tr A_{-n;ab}^\alpha \delta Q  , \notag \\
O_t^{(2),2} &=&- \frac{\pi T}{4} \int d \bm{r}\sum_{\alpha n; a b} \bm{\Gamma}_{ab} \tr
A_{n;ab}^\alpha \delta Q \tr A_{-n;ab}^\alpha \delta Q  , \notag\\
O_\eta &=& 4\pi T z \int d \bm{r}\tr A_\eta \delta Q .
\end{eqnarray}
Here we introduce $\delta Q = Q-\Lambda$ and
\begin{equation}
A_\eta = T_0 [\eta, T_0^{-1}], \quad A_{n;ab}^\alpha = T_0 [ I_n^\alpha t_{ab}, T_0^{-1}] .
\end{equation}
The effective action $\mathcal{S}_{\rm eff}[Q_{0}]$ can be obtained by expansion of $\mathcal{S}[T_0^{-1} Q T_{0}]$ to the second order in $A_\eta$ and $A_{n;ab}^\alpha$.~\cite{BPS} Then, we find
\begin{gather}
\mathcal{S}_{\rm eff}[Q_{0}] - \mathcal{S}_F[Q_0] = \langle O_t^{(2),1} \rangle + \langle O_t^{(2),2} \rangle  + \frac{1}{2}  \langle \bigl [O_t^{(1),1}\bigr ]^2 \rangle\notag \\
+ \frac{1}{2}  \langle O_t^{(1),1} O_t^{(1),2} \rangle + \frac{1}{2}  \langle \bigl [O_t^{(1),2}\bigr ]^2 \rangle+ \langle O_\eta \rangle , \label{SeffA2}
\end{gather}
where the average $\langle\dots\rangle$ is with respect to action~\eqref{SstartSigma}-\eqref{SstartF} and we omit terms which do not involve infrared divergencies. In general, each term in the right hand side of Eq. \eqref{SeffA2} produce contributions which cannot be expressed in terms of $Q_0$ only.  However, all such contributions cancel in the total expression~\eqref{SeffA2}. Therefore, we will not list them below.  Expanding $\delta Q$ in series of W according to Eq.~\eqref{Qexp} and performing averaging with the help of Eq.~\eqref{Prop}, we obtain
\begin{gather}
\mathcal{S}_{\rm eff}[Q_{0}] = - \frac{\pi T}{4} \int d \bm{r}\sum_{\alpha n; a b} {\Gamma}^\prime_{ab} \tr
I_n^\alpha t_{ab} Q \tr I_{-n}^\alpha t_{ab} Q \notag \\
+4\pi T z^\prime  \int d \bm{r}\tr \eta Q 
\end{gather}
where
\begin{gather}
{\Gamma}^\prime_{ab} = {\Gamma}_{ab} +\delta {\Gamma}^{(2),1}_{ab}+\delta {\Gamma}^{(2),2}_{ab}+\delta {\Gamma}^{(1),1;1}_{ab}+\delta {\Gamma}^{(1),1;2}_{ab}\notag \\
+\delta {\Gamma}^{(1),2;2}_{ab}+\delta {\Gamma}^{\eta}_{ab} \label{dG1}
\end{gather}
and similar for $z^\prime$. Here the contributions to $\Gamma_{ab}^\prime$  from each term in the right hand side of Eq.~\eqref{SeffA2} are given as follows
\begin{gather}
\langle O_t^{(2),1} \rangle \to \delta {\Gamma}^{(2),1}_{ab} = \frac{32\pi T}{\sigma_{xx}^2} \sum_{cd;ef} \bigl [\mathcal{C}_{cd;ef}^{ab} \bigr ]^2 \Gamma_{cd}\Gamma_{ef}  \notag \\
\times  \int \frac{d^2 \bm{p}}{(2\pi)^2}\, \sum_{\omega_m>0} DD_p^{(cd)}(\omega_m) ,
\end{gather}
\begin{gather}
\langle O_t^{(2),2} \rangle \to \delta {\Gamma}^{(2),2}_{ab} = -\frac{1}{8\sigma_{xx}} \sum_{cd;ef} \bigl [\Sp (t_{cd}t_{ef} t_{ab}) \bigr ]^2 \Gamma_{cd} \notag \\
\times \int \frac{d^2 \bm{p}}{(2\pi)^2}\, D_p(0) ,
\end{gather}
\begin{gather}
\frac{1}{2}\langle \bigl [ O_t^{(1),1}\bigr ]^2 \rangle \to \delta {\Gamma}^{(1),1;1}_{ab} = \frac{32\pi T}{\sigma_{xx}^2} \sum_{cd;ef} \bigl [\Gamma_{ab} \mathcal{C}_{cd;ef}^{ab} \bigr ]^2 \sum_{\omega_m>0}\notag \\
\times   \int \frac{d^2 \bm{p}}{(2\pi)^2} \bigl [ D^{(cd)}D_p^{(ef)}(\omega_m)-D^2_p(\omega_m)\bigr ] ,
\end{gather}
\begin{gather}
\langle O_t^{(1),1}O_t^{(1),2} \rangle \to \delta {\Gamma}^{(1),1;2}_{ab} = -\frac{64\pi T}{\sigma_{xx}^2} \sum_{cd;ef} \bigl [\mathcal{C}_{cd;ef}^{ab} \bigr ]^2 \Gamma_{ab}  \Gamma_{cd} \notag \\
\times   \int \frac{d^2 \bm{p}}{(2\pi)^2} \,\sum_{\omega_m>0} D^{(ab)}D_p^{(ef)}(\omega_m) ,
\end{gather}
\begin{gather}
\frac{1}{2}\langle \bigl [ O_t^{(1),2}\bigr ]^2 \rangle \to \delta {\Gamma}^{(1),2;2}_{ab} = \frac{32\pi T}{\sigma_{xx}^2} \sum_{cd;ef} \bigl [\Gamma_{ef} \mathcal{C}_{cd;ef}^{ab} \bigr ]^2 \sum_{\omega_m>0}\notag \\
\times   \int \frac{d^2 \bm{p}}{(2\pi)^2} \bigl [D^{(cd)}D^{(ef)}_p(\omega_m)-DD_p^{(cd)}(\omega_m)\bigr ] ,
\end{gather}
and 
\begin{gather}
\langle O_\eta \rangle \to \delta {\Gamma}^{\eta}_{ab} = 0 .\label{dG2}
\end{gather}
Combing contributions~\eqref{dG1}-\eqref{dG2} we obtain Eq.~\eqref{RenGam1}. 

The only non-zero contributions to renormalization of $z$ are 
\begin{gather}
\frac{1}{2}\langle \bigl [ O_t^{(1),2}\bigr ]^2 \rangle \to \delta z^{(1),2;2}= \frac{64\pi T}{\sigma_{xx}^2} \sum_{cd} \Gamma_{cd}  \sum_{\omega_m>0}\notag \\
\times   \int \frac{d^2 \bm{p}}{(2\pi)^2} \bigl [DD_p(\omega_m)-D^{(cd)}D_p^{(cd)}(\omega_m)\bigr ] \label{dz11}
\end{gather}
and
\begin{gather}
\langle O_\eta \rangle \to \delta z^{\eta} = -\frac{64\pi T}{\sigma_{xx}^2} \sum_{cd} \Gamma_{cd}\int \frac{d^2 \bm{p}}{(2\pi)^2}\notag \\
\times \sum_{\omega_m>0} DD_p^{(cd)}(\omega_m) . \label{dz2}
\end{gather}
In total, Eqs.~\eqref{dz11} and \eqref{dz2} give
\begin{gather}
z^\prime=z+ \frac{64\pi T}{\sigma_{xx}^2} \sum_{cd} \Gamma_{cd} \int \frac{d^2 \bm{p}}{(2\pi)^2}\,  \sum_{\omega_m>0} D^2_p(\omega_m) . \label{dz1}
\end{gather}
It coincides with Eq.~\eqref{z2}.

\section{Evaluation of DC �transconductance $\sigma_{12}^\prime$ \label{TC}}

In this appendix we present calculations of the DC transconductance in the one-loop approximation.
Similarly to the total conductance, the transconductance can be obtained from
\begin{gather}
\sigma^\prime_{12}(i\omega_n) = -\frac{\sigma_{xx}}{16 n}\left
\langle\tr[I_{n}^\alpha t_-, Q][I_{-n}^\alpha t_+, Q] \right
\rangle + \frac{\sigma_{xx}^2}{128 n } \int
d\bm{r}^\prime \notag 
\\
\times \langle\langle \tr I_n^\alpha t_-
Q(\bm{r})\nabla Q(\bm{r}) \tr I_{-n}^\alpha t_+
Q(\bm{r}^\prime)\nabla Q(\bm{r}^\prime)\rangle \rangle
\label{TCODef}
\end{gather}
after the analytic continuation to the real frequencies: $i\omega_n \to
\omega + i0^+$ at $\omega \to 0$. Here matrices $t_\pm = (t_{00}\pm t_{30})/2$. 
Evaluation of the transconductance according to Eq.~\eqref{TCODef} in the one-loop
approximation yields
\begin{gather}
\sigma^\prime_{12}(i\omega_n) = \frac{32 \pi T}{\sigma_{xx} \omega_n} \sum_{ab} \Gamma_{ab} \Sp\bigl [ t_- t_{ab} t_+ t_{ab} - t_- t_+ \bigr ] \sum_{\omega_m>0}\notag \\
\times  \omega_m \int \frac{d^2 \bm{p}}{(2\pi)^2}\,  DD^{(ab)}_p(\omega_{m+n})
\notag \\
-\frac{8 \pi T}{\sigma_{xx} \omega_n}\int \frac{d^2 \bm{p}}{(2\pi)^2}\, p^2  \sum_{ab;cd}  \Sp \bigl [ t_- t_{ab} t_{cd}\bigr ]
\sum_{\omega_m>0}  \notag \\
\times\Biggl \{ \Sp \bigl ( t_+ [ t_{ab}, t_{cd} ]\bigr ) \omega_m \Bigl [ \Gamma_{ab} D_p(\omega_{m+n})DD^{(ab)}_p(\omega_m) \notag \\
+ \Gamma_{cd} D_p^{(ab)}(\omega_m)DD^{(cd)}_p(\omega_{m+n}) \Bigr ]
- \Sp \bigl [ t_+t_{ab} t_{cd} \bigr ] \notag \\
\times \Gamma_{ab} \min \{\omega_m,\omega_n\} D_p(\omega_{m+n})DD^{(ab)}_p(\omega_m) \Biggr \},
\label{SRTC}
\end{gather}
where $DD^{(ab)}_p(\omega_m)\equiv D_p(\omega_m) D^{(ab)}_p(\omega_m)$. Evaluating the traces we find
\begin{gather}
\sigma^\prime_{12}(i\omega_n) = \frac{2^9 \pi T\Gamma_v}{\sigma_{xx} \omega_n}  \int \frac{d^2 \bm{p}}{(2\pi)^2}\, \sum_{\omega_m>0} \Biggl \{ \omega_m  \Bigl [ DD^{(20)}_p(\omega_{m+n})\notag \\
 - p^2 \Bigl ( D_p(\omega_m) +D^{(20)}_p(\omega_{m+n}) \Bigr ) D_p(\omega_{m+n})D^{(20)}_p(\omega_m) \Bigr ] \notag \\
+ \min \{\omega_m,\omega_n\} p^2 D_p(\omega_{m+n})DD^{(20)}_p(\omega_m) \Biggr \} .
\label{SRTC1}
\end{gather}
Performing the analytic continuation to the real frequencies, $i\omega_n\to\omega+i0^+$, one obtains the DC
transconductance in the one-loop approximation:
\begin{gather}
\sigma^\prime_{12} = - \frac{2^7 \Gamma_v}{\sigma_{xx}}\Im  \int \frac{d^2 \bm{p}}{(2\pi)^2}\, \int d\Omega \Biggl \{  
\frac{\partial}{\partial\Omega} \left (\Omega \coth \frac{\Omega}{2T}\right ) \notag \\
\times \Bigl [  DD^{(20),R}_p(\Omega) -  p^2 D^2D^{(20),R}_p(\Omega) \Bigr ] +  p^2 \Omega \coth \frac{\Omega}{2T}   \notag \\
\times D^{(20),R}_p(\Omega) \frac{\partial}{\partial\Omega}  \Bigl [ \frac{1}{2}D^2_p(\Omega) + DD^{(20),R}_p(\Omega) \Bigr ] \Biggr \} ,
\label{SRTC2}
\end{gather}
where $DD^{(ab),R}_p(\Omega)\equiv D^R_p(\Omega) D^{(ab),R}_p(\Omega)$. Next, Eq.~\eqref{SRTC2} can be simplified as
\begin{gather}
\sigma^\prime_{12} = \frac{2^{11} \Gamma_v}{\sigma^2_{xx}}\Re  \int d\Omega \,\Omega \coth\frac{\Omega}{2T} 
\notag \\
\times
\int \frac{d^2 \bm{p}}{(2\pi)^2}\,p^2
 DD^{(20),R}_p(\Omega)
\notag \\
\times \Biggl \{
z  \Bigl [ D_p^{R}(\Omega) \Bigr ]^2 -(z+\Gamma_v) \Bigl [ D_p^{(20),R}(\Omega) \Bigr ]^2
\Biggr \} 
\end{gather}
One can check that due to integration over momentum $p$ the DC transconductance vanishes at arbitrary temperature, $\sigma_{12}^\prime=0$.

%%%%%%%%%%%%%%%%%%%%%%%%%%%%%%%%%%


\begin{thebibliography}{100}
\bibitem{Footnote} In the experiments of Refs.~[\protect\onlinecite{MinkovGermanenko1,MinkovGermanenko2}] the random potential was created by charged impurities situated near $z=0$. In this case the range of $W$ is determined by 3D screening length $d_W \sim 1/\sqrt{\varkappa k_F}$. In order to consider this random potential as short-ranged the following condition $d_W\ll l$ or, equivalently, $k_F l \gg \sqrt{k_F/\varkappa}$ should hold.
\end{thebibliography}
\end{document}